\documentclass[]{aastex631}

\usepackage{graphicx}	% Including figure files
\usepackage{amsmath}	% Advanced maths commands
\usepackage{amssymb}	% Extra maths symbols
\usepackage{xcolor}     % Nice colours
\usepackage{tabularx}

\usepackage{ulem}

\newcommand\aastex{AAS\TeX}
\newcommand\latex{La\TeX}

\newcommand{\Msun} {M$_{\odot}$}

\begin{document}

\title{{\it LISA} Galactic binaries with astrometry from Gaia DR3}

\author[0000-0002-6540-1484]{Thomas Kupfer}
\affiliation{Hamburger Sternwarte, University of Hamburg, Gojenbergsweg 112, 21029 Hamburg, Germany}
\affiliation{Department of Physics and Astronomy, Texas Tech University, PO Box 41051, Lubbock, TX 79409, USA}

\author[0000-0002-6725-5935]{Valeriya Korol}
\affiliation{Max-Planck-Institut f{\"u}r Astrophysik, Karl-Schwarzschild-Straße 1, 85741 Garching, Germany}
\affiliation{Institute for Gravitational Wave Astronomy, School of Physics and Astronomy, University of Birmingham, Birmingham, B15 2TT, UK}

\collaboration{20}{both authors contributed equally to this work}

\author{Tyson B. Littenberg}
\affiliation{NASA Marshall Space Flight Center, Huntsville, Alabama 35811, USA}

\author[0000-0002-0716-1320]{Sweta Shah}
\affiliation{Max Planck Institute for Gravitational Physics (Albert Einstein Institute), Callinstrasse 38, 30167 Hannover, Germany} 
\affiliation{Leibniz Universit\"at Hannover, Institut für Gravitationsphysik, Callinstrasse 38, 30167 Hannover, Germany}

\author[0000-0001-6201-0560]{Etienne Savalle}
\affiliation{Universit\'e de Paris, CNRS, Astroparticule et Cosmologie, 75013 Paris, France}
\affiliation{IRFU, CEA, Universit\'e Paris-Saclay, F-91191, Gif-sur-Yvette, France}

\author{Paul J. Groot}
\affiliation{Department of Astrophysics/IMAPP, Radboud University, P.O.Box 9010, 6500 GL Nijmegen, The Netherlands}
\affiliation{South African Astronomical Observatory, PO Box 9, Observatory, 7935, Cape Town, South Africa}
\affiliation{Department of Astronomy \& Inter-University Institute for Data Intensive Astronomy, University of Cape Town, Private Bag X3, 7701 Rondebosch, South Africa}

\author[0000-0002-2498-7589]{Thomas R. Marsh}
\affiliation{Department of Physics, University of Warwick, Gibbet Hill Road, Coventry CV4 7AL, UK}

\author{Maude Le Jeune}
\affiliation{Universit\'e de Paris, CNRS, Astroparticule et Cosmologie, 75013 Paris, France}

\author{Gijs Nelemans}
\affiliation{Department of Astrophysics/IMAPP, Radboud University, P.O.Box 9010, 6500 GL Nijmegen, The Netherlands}
\affiliation{SRON, Netherlands Institute for Space Research, Niels Bohrweg 4, 2333 CA Leiden, The Netherlands}
\affiliation{Institute of Astronomy, KU Leuven, Celestijnenlaan 200D, B-3001 Leuven, Belgium}

\author[0000-0001-7069-7403]{Anna F. Pala}
\affiliation{European Space Agency, European Space Astronomy Centre, Camino Bajo del Castillo s/n, 28692 Villanueva de la Cañada, Madrid, Spain}

\author[0000-0002-7371-9695]{Antoine Petiteau}
\affiliation{Universit\'e de Paris, CNRS, Astroparticule et Cosmologie, 75013 Paris, France}
\affiliation{IRFU, CEA, Universit\'e Paris-Saclay, F-91191, Gif-sur-Yvette, France}

\author[0000-0001-8722-9710]{Gavin Ramsay}
\affiliation{Armagh Observatory and Planetarium, College Hill, Armagh BT61 9DG, UK}

\author[0000-0003-0771-4746]{Danny Steeghs}
\affiliation{Department of Physics, University of Warwick, Gibbet Hill Road, Coventry CV4 7AL, UK}

%\author[0000-0002-0587-6680]{Elena Maria Rossi}
%\affiliation{Leiden Observatory, Leiden University, PO Box 9513, NL-2300 RA Leiden, The Netherlands}

\author{Stanislav Babak}
\affiliation{Universit\'e de Paris, CNRS, Astroparticule et Cosmologie, 75013 Paris, France}

\begin{abstract}
Galactic compact binaries with orbital periods shorter than a few hours emit detectable gravitational waves at low frequencies. Their gravitational wave signals can be detected with the future {\it Laser Interferometer Space Antenna} ({\it LISA}).  Crucially, they may be useful in the early months of the mission operation in helping to validate {\it LISA}'s performance in comparison to pre-launch expectations. We present an updated list of 55 candidate {\it LISA} detectable binaries with measured properties, for which we derive distances based on {\it Gaia} Data release 3 astrometry. Based on the known properties from EM observations, we predict the {\it LISA} detectability  after 1, 3, 6, and 48 months using Bayesian analysis methods. We distinguish between verification and detectable binaries as being detectable after 3 and 48\, months respectively. We find 18 verification binaries and 22 detectable sources, which triples the number of known {\it LISA} binaries over the last few years. These include detached double white dwarfs, AM\,CVn binaries, one ultracompact X-ray binary and two hot subdwarf binaries. We find that across this sample the gravitational wave amplitude is expected to be measured to $\approx10\%$ on average, while the inclination is expected to be determined with $\approx15^\circ$ precision. For detectable binaries these average errors increase to $\approx50\%$ and to $\approx40^\circ$ respectively. 

%Until a few years ago about one dozen binaries were known with properties making them detectable for {\it LISA}. 
%We find that the parallax precision has improved from {\it Gaia} Data release 2 to early {\it Gaia} Data release 3 by $20 - 30$\%. 
%We distinguish between {\it LISA} verification binaries, which are detected after only 3 months of {\it LISA} observations and {\it LISA} detectable binaries, which are detected after 48 months of {\it LISA} observations. 

\end{abstract}

%Classical Novae (251) --- Ultraviolet astronomy(1736) --- History of astronomy(1868) --- Interdisciplinary astronomy(804)}


\section{Introduction} \label{sec:intro}

Binary systems composed of degenerate stellar remnants (white dwarfs, neutron stars and black holes) in orbits with periods of less than a few hours are predicted to be strong gravitational wave (GW) sources in our own Galaxy. 
%\sout{Perhaps the most prominent example of such a system is the first binary pulsar discovered by \citet{HT1975}, which via the measurement of its orbital decay over three decades, led to the first indirect evidence for the existence of GWs \citep{Weisberg:2005} and paved the way forward for the field of GW astronomy. Since then,} 
A number of these systems -- primarily consisting of a neutron star or white dwarf paired with a compact helium-star, white dwarf, or another neutron star -- have been identified primarily through the observation in optical and X-ray electromagnetic (EM) wavebands. Some of these systems display remarkably short orbital periods, down to just several minutes \citep[e.g.][]{LISAastroWP}.
%Many of such systems consisting of a neutron star/white dwarf with a compact helium-star/white dwarf/neutron star companion with orbital periods as short as several minutes have been discovered, mainly at optical or X-ray electromagnetic (EM) wavebands \citep[e.g.][]{LISAastroWP}. 
Binaries in a such a tight orbit emit GWs at mHz frequencies that can be detected directly with the future {\it Laser Interferometer Space Antenna} \citep[{\it LISA},][]{ama17}, and other future planned space-based GW observatories such as \textit{TianQin} \citep{TianQin, hua20}, \textit{Taiji} \citep{Taiji} and the \textit{Lunar Gravitational Wave Antenna} \citep{LGWA}. 

In this study we focus on the {\it LISA} mission, which is a European Space Agency (ESA)-led GW observatory currently scheduled for launch in the mid-2030s\footnote{https://sci.esa.int/web/lisa/-/61367-mission-summary}. Designed to operate in the frequency band between $0.1$\,mHz and $100$\,mHz \citep{ama17}, {\it LISA} is an ideal tool for discovering massive black hole mergers and extreme-/intermediate-mass ratio inspirals. In addition, it can survey the shortest period stellar remnant binaries throughout the entire Milky Way, providing a complementary view of our Galaxy to EM surveys \citep[for a review see][]{LISAastroWP}. Both theory- and observation-based studies find that {\it LISA} will deliver a sample of ${\cal O} (10^4)$ binaries with orbital periods of $< 1$\,hour, which will be complete up to periods of $< 15\,$min \citep[e.g.][]{nel01,rui09,nis12,lam19,bre20,li20,kor22}. A significant number of ${\cal O} (10^2)$ of stellar remnant binaries -- primarily those composed of two white dwarfs -- discovered by {\it LISA} will be possible to study in combination with EM surveys \citep[e.g.][]{nel04,mar11,kor17,bre18,tau18,li20}.

In the context of the {\it LISA} mission, stellar remnant binaries known from EM observations are often termed `verification binaries', based on the idea that one can model their GW signal using EM measurements of binary's parameters and to employ these to test {\it LISA} data quality \citep[e.g.][]{str06,lit18timing,sav22}. In our previous work, we reviewed a sample of candidate {\it LISA} verification binaries following the second {\it Gaia} data release (DR2). This allowed us to determine distances -- previously highly uncertain for most binaries -- based on {\it Gaia}'s parallax measurements \citep{kup18,ram18}. In turn, new distance estimates enabled us to evaluate the uncertainty on the expected GW signal's amplitude and to assess the detectability of these candidate verification binaries with {\it LISA}. In this work we update the sample of candidates {\it LISA} verification binaries in a number of ways. Firstly, we include several newly discovered systems since {\it Gaia} DR2 (Section~\ref{sec:sample}). Secondly, we re-evaluate the distances based on improved astrometry from the third {\it Gaia} DR3, while also taking into account their proper motion information (Section~\ref{sec:distance}). In addition, we evaluate their detectability as well as the binary parameter estimation in a fully-Bayesian way using the up-to-date {\it LISA} sensitivity requirements (Section~\ref{sec:ldsoft}).%\citep{LISAdoc}.  

So far verification binaries have been (arbitrarily) defined as such based on an assumed signal-to-noise ratio (SNR) detection threshold reached at a set observation time. However, this definition relies on a few caveats. Firstly, the SNR threshold and the integration time needed to make an unambiguous identification of a (known) source is not just a matter of source's intrinsic amplitude, but is also heavily dependent on the realization of the rest of the Galactic population (i.e. unresolved Galactic confusion foreground) that the source is competing against \citep[e.g. figure 4 of ][]{kor23}. In addition, the Galactic confusion foreground is dynamic: it will decrease with time as more and more sources will become detectable and will become individually resolved. Moreover, {\it LISA}'s orbit around the Sun will introduce a modulation in the Galactic foreground reaching its maximum when {\it LISA} is optimally oriented towards the Galactic center, which is where the density of Galactic sources peaks \citep[e.g.][]{petiteauPhDthesis}. A prototype global fit data analysis pipeline for {\it LISA} demonstrated that the Galactic foreground subtraction steadily improves with time with a few ${\cal O}(10^3)$ binaries being identified (and subtracted) already after 1 month \citep{lit23}. Moreover, known binaries will be the most crucial in the early weeks/months of the mission operation in helping to validate the early performance of the instrument in comparison to pre-launch expectations. It is therefore reasonable to expect that the first data validation may be required after only a few months from the beginning of science operations. %, e.g. to guarantee the first data release within six months - one year of mission. 
%We note that the data release scenario for {\it LISA} is still to be defined by ESA. <- Is no longher true
We anticipate that an integration time as short as 1-3 months would allow for basic consistency tests on the recovered parameters on a few epochs of commissioning data for several verification binaries. Given all the above, in this study we opt to call as {\it verification binary} a system that becomes detectable, which is judged based on the shape of the recovered posteriors on binary's parameters rather than a SNR threshold, within 3 months of observation time with {\it LISA}, and we call a as {\it detectable binary} when it is detected after 48 months (at present set as the nominal lifetime of the mission).

\section{The sample of compact {\it LISA} sources since {\it Gaia} DR2} \label{sec:sample}
At present the catalog of candidate verification binaries includes detached \citep{bro16a} and semi-detached double white dwarfs  (the latter called AM\,CVn type binaries; see \citet{sol10} for a recent review), hot subdwarf stars with a white dwarf companion (see \citet{gei13, kup22, pel21} for recent discoveries), semi-detached white dwarf-neutron star binaries (so-called ultracompact X-ray binaries; \citealt{nel10a}), double neutron stars \citep{lyn04} and Cataclysmic Variables \citep[CVs,][]{sca23}.
In \citet{kup18} we analyzed $\sim 50$ known candidates using distances derived from parallaxes provided in the {\it Gaia} DR2 catalog \citep[DR2,][]{gai18}. We found that 13 candidates exceed a SNR threshold of 5 for a {\it LISA} mission duration of 4 years.

The Zwicky Transient Facility (ZTF) performed a dedicated high-cadence survey to find short period binaries \citep{kup21}. More than 20 new binary systems with orbital periods ranging from 7\,min to $\approx1$\,h have been discovered by ZTF since the beginning of science operations in March 2018 \citep{bel14,bur19,bur20,bur20a,kup20,kup20a,roe22,bur23}. This new sample includes eight eclipsing systems, seven AM\,CVn systems, and six systems exhibiting primarily ellipsoidal variations in their light curves. Remarkably, one of the first ZTF discoveries was the shortest orbital period detached eclipsing binary system known to date, ZTF J1539+5027, with an period of just 6.91 min \citep{bur19}. Owing both to its inherently high GW frequency and large GW amplitude, ZTF J1539+5027 is expected to be one of the loudest Galactic GW sources and could reach the SNR detection threshold of $\approx7$ within a week. \citet{lit19} showed that for high frequency systems like ZTF J1539+5027, GW measurements will independently provide comparable levels of precision to the current EM measurement of the orbital evolution of the system, and will improve the precision to which the distance and orientation is known. 

\begin{figure}
  \begin{center}
    \includegraphics[width=0.6\textwidth]{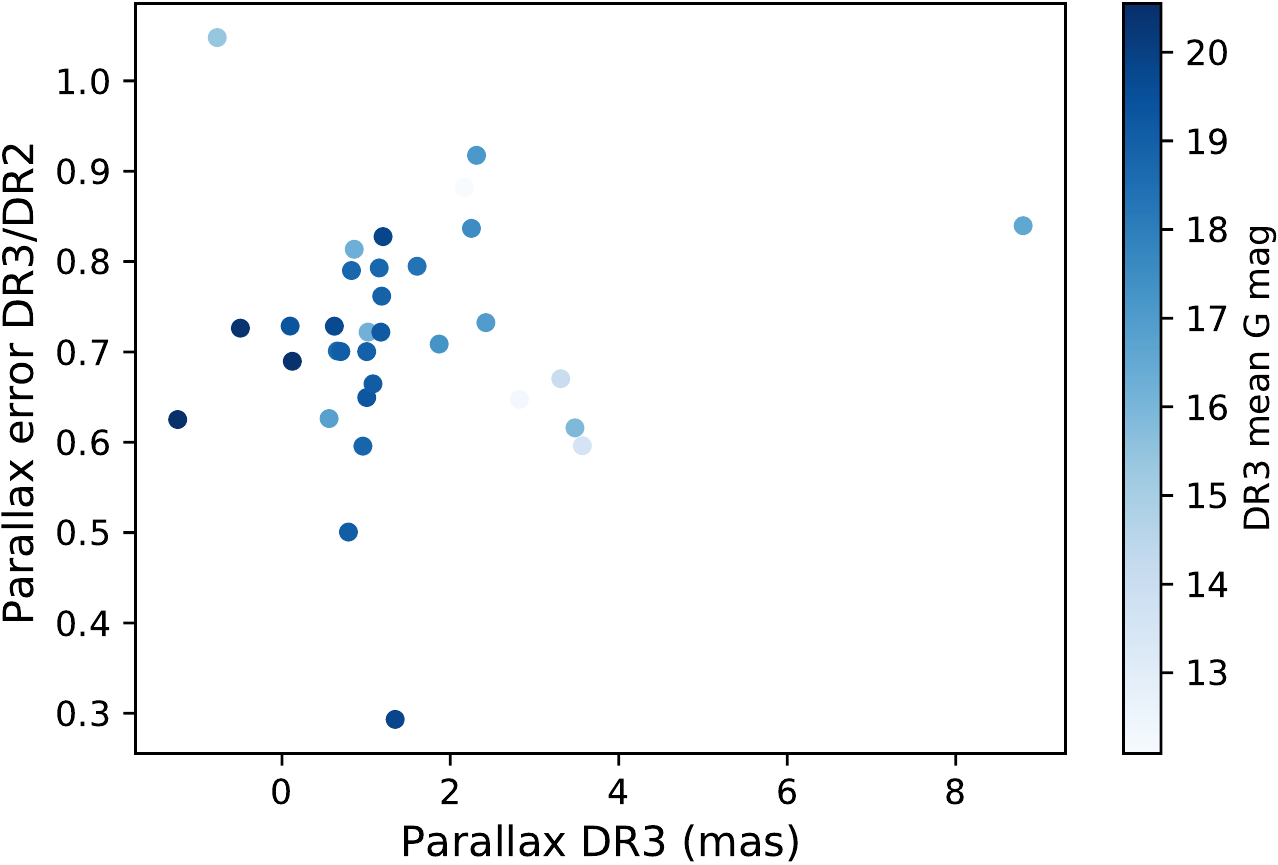} 
\vspace{2mm}
  \caption{The error of the parallax in DR3 compared to DR2 for the verification sources in Tables 1 and 2. The color indicates the mean $G$ mag of the source. }
    \label{fig:parallaxerrDR2vEDR3}
    \end{center}
\end{figure}

The Extremely Low Mass (ELM) White Dwarf Survey has successfully completed its observations across the footprint of the Sloan Digital Sky Survey (SDSS), and expanded their search to the Southern hemisphere \citep{bro22,kos20,kos23a}. Over the last few years the ELM survey discovered several sub-hour orbital-period double degenerates, including the first double helium-core white dwarf binary (e.g. \citealt{bro20,kil21,kos21,kos23}). It is expected that double helium-core white dwarfs and carbon/oxygen + helium-core white dwarfs dominate the population in the {\it LISA} band despite that they make up only 10\% of the global double white dwarf population \citep{lam19}.

Moreover, several additional candidates have been found in other large-scale surveys. SDSS\,J1337 was discovered as a double degenerate in early SDSS-V data with an orbital period of 99\,min. The spectrum shows spectral lines from both components making it a double lined system which provides precise system parameters \citep{cha21}. \citet{pel21} discovered a compact hot subdwarf binary with a massive white dwarf companion in a 99\,min period orbit in the TESS sky survey. The total mass of the system is above the Chandrasekhar mass making the system a double degenerate supernova Ia progenitor. \citet{sca23} showed that three known CVs, namely WZ\,Sge, VW\,Hyi and EX\,Hya, could be individually resolved after four years of {\it LISA} operation.

Since the release of {\it Gaia} DR2 a few years ago, numerous sky surveys have collectively tripled the number of identified candidate compact binaries. It's noteworthy, however, that these surveys employ a variety of detection techniques and analysis methods, leading to heterogeneity in the presentation of results. Table \ref{tab:system} offers an overview of the observational results for known sources detectable by {\it LISA}, as reported in the respective studies. The properties of all sources compiled for this publication are accessible to the public via the {\it LISA} Consortium's GitLab repository at \url{https://gitlab.in2p3.fr/LISA/lisa-verification-binaries}.

\begin{table}
\begin{center}
\caption{Physical properties of the known verification and detectable binaries. Masses and inclination angles in brackets are assumed and based on evolutionary stage and mass ratio estimations. Absolute magnitudes are calculated from the {\it Gaia} $G$-band magnitude in combination with our distance estimate.} 
\begin{tabularx}{\textwidth}{lllrrrrlll}
\hline
Source       &  Type   & Orbital      &  $l_{\rm Gal}$   &  $b_{\rm Gal}$  &   BP-RP    &   $M_\textrm{G}$    &  $m_{\rm 1}$ & $m_{\rm 2}$ & $\iota$   \\
             &          & period (s)     &   (deg)  & (deg)      &   (mag)  &    (mag)    & (\Msun) &    (\Msun)  &  ($\deg$)    \\
\hline
 \multicolumn{3}{l}{{\bf Verification binaries}}      &     \\
HM Cnc$^{1,2}$       &  AMCVn     & 321.529129(10) &  206.9246  &   23.3952  & $0.246$  & 6.54  &    0.55     &  0.27  & $\approx$38  \\
ZTFJ1539$^{3}$    &  DWD$^*$   &  414.7915404(29) &  80.7746  &   50.5819  &  $-0.263$ &  8.44 &  $0.61^{+0.017}_{-0.022}$  & $0.21\pm0.015$  &  $84.15^{+0.64}_{-0.57}$  \\
ZTFJ2243$^{4}$    &  DWD$^*$  &  527.934890(32)  &   104.1514	& --5.4496   & $-0.160$  &  9.33 &  $0.349^{+0.093}_{-0.074}$  &  $0.384^{+0.114}_{-0.074}$  &  $81.88^{+1.31}_{-0.69}$ \\
V407 Vul$^5$     &  AMCVn     & 569.396230(126) & 57.7281  &    6.4006  & $1.535$  & 7.76  &   [0.8$\pm$0.1]   & [0.177$\pm$0.071]   & [60]  \\
ES Cet$^{6}$       &  AMCVn$^*$ &   620.21125(30) &   168.9684  & --65.8632  & $-0.296$ &  5.55 &  [0.8$\pm$0.1]  &  [0.161$\pm$0.064] &  [60] \\
SDSSJ0651$^{*7,8}$  &  DWD$^*$  &  765.206543(55)  &  186.9277  &   12.6886  & $0.029$ & 9.37  &   0.247$\pm$0.015 &  0.49$\pm$0.02 & ${86.9^{+1.6}_{-1.0}}$ \\ ZTFJ0538$^{9}$     &  DWD$^*$  &  866.60331(16)  &  186.8104	&  --6.2213  & $0.025$ & 8.80  &  $0.45\pm0.05$  &  $0.32\pm0.03$  &  $85.43^{+0.07}_{-0.09}$ \\   
SDSSJ1351$^{10}$  &  AMCVn     &  939.0(7.2)  &  328.5021  &   53.1240  &  $-0.122$ & 7.80  &     [0.8$\pm$0.1]     &  [0.100$\pm$0.040]  & [60]  \\
AM CVn$^{11,12}$        &  AMCVn    &  1028.7322(3)  & 140.2343  &   78.9382  & $-0.283$ &  6.66 &    0.68$\pm$0.06 &  0.125$\pm$0.012  & 43$\pm$2  \\
ZTFJ1905$^{9}$  & AMCVn$^*$   &  1032.16441(62)   & 0.1945  &  1.0968  & $-0.066$  & 11.47  &  [0.8$\pm$0.1]  &  [0.090$\pm$0.035]  &  $70\pm20$    \\
SDSSJ1908$^{13,14}$   &  AMCVn    &  1085.108(1)  &  70.6664  &   13.9349    & $-0.018$  & 6.27 & [0.8$\pm$0.1]  & [0.085$\pm$0.034]  & 10 - 20  \\
HP Lib$^{15,16}$        &  AMCVn    & 1102.70(5)   &  352.0561  &   32.5467    & $-0.153$ &  6.36 &  0.49-0.80 & 0.048-0.088 & 26-34  \\
SDSSJ0935$^{17,18}$   &  DWD   & 1188(42)  &  176.0796  &   47.3776    & $0.455$  & 9.82  &  0.312$\pm$0.019   &   0.75$\pm$0.24 &  [60] \\
J0526+5934$^{19}$  &   DWD   & 1230.37467(7) & 151.9201  & 13.2614  &  $0.186$  &  7.92  &   $0.378^{+0.066}_{-0.060}$   &    $0.887^{+0.110}_{-0.098}$   & $57.1^{+4.3}_{-4.1}$  \\
J1239-2041$^{20}$  & DWD   &  1350.432(11.232) &  299.2755 & 42.0943  & $-0.072$  & 9.00  &  $0.291\pm0.013$  & ${0.68^{+0.11}_{-0.06}}$   &  ${71^{+8}_{-10}}$     \\
TIC378898110       & AMCVn  &  1347.96   &   297.0555   &  1.9451  &  0.027   &  6.82  &   [0.8$\pm$0.1] &  [0.10$\pm$0.02]  &  $74\pm10$  \\
CR Boo$^{16,21}$        &  AMCVn  & 1471.3056(500) &  340.9671  &   66.4884    & $0.066$ &  7.74 &   0.67-1.10 &  0.044-0.088 & 30  \\
SDSS0634$^{22}$  &  DWD  &   1591.4(28.9)  & 176.7322  &  13.3211  &  $-0.189$  & 8.95  &  ${0.452^{+0.070}_{-0.062}}$  &  ${0.209^{+0.034}_{-0.021}}$  &  $37\pm7$    \\
V803 Cen$^{16,23}$      &  AMCVn   &  1596.4(1.2)     &  309.3671  &   20.7262   & $0.232$ & 8.44  &  0.78-1.17 & 0.059-0.109 & 12 - 15  \\

%ASASSN-14cc              & 22.5 (sh) &  &  &  & & & &\\

%KL Dra                     &   91.0140  &   19.1992  &  1500.0   & 0.76 & 0.057 & [60] & 13,14 \\
%PTF1J2219+3135           & 26.1 & &  &   & & & & & \\

%PTF1 J071912.13+485834.0   &  168.6573  &   24.4945  &  1606.2   &  [0.8$\pm$0.1] & [0.053$\pm$0.021] & [60] &   \\
%SDSS J092638.71+362402.4   &  187.5084  &   46.0110  &  1698.6   &  0.85$\pm$0.04 &  0.035$\pm$0.003 & 82.6$\pm$0.3 & 16 \\
%CP Eri                     &  191.7017  & --52.9098  &  1704.0   &  [0.8$\pm$0.1] & [0.049$\pm$0.020] & [60] & \\
%V406 Hya                   &  235.1276  &   26.4800  &  2027.8   &  [0.8$\pm$0.1] & [0.040$\pm$0.016] &  [60] &  \\
%SDSS J124058.03-015919.2   &  297.5679  &   60.7740  &  2241.3   &  [0.8$\pm$0.1] & [0.035$\pm$0.014] &  [60] &  \\
%SDSS J012940.05+384210.4   &  131.0586  & --23.5648  &  2253.2   &  [0.8$\pm$0.1] & [0.034$\pm$0.014] &  [60] &  \\
%SDSS J080449.49+161624.8   &  205.9478  &   23.3716  &  2670.0   &  [0.8$\pm$0.1] & [0.027$\pm$0.011] &  [60] &  \\
%GP Com                     &  323.5224  &   80.3141  &  2791.2   &  0.50-0.68  & 0.010-0.012 & [60] & 9,17 \\
%{\it Gaia} 14aae                 &   95.3090  &   41.6879  &  2982.49  &  0.87$\pm$0.02 & 0.0250$\pm$0.0013 &  86.3$\pm$0.3  &  \\
%SDSS J120841.96+355025.1   &  166.5229  &   77.4097  &  3177.6   &   [0.8$\pm$0.1] & [0.022$\pm$0.009] & [60] & \\
%V396 Hya                   &  309.2593  &   39.2506  &  3906.0   &   [0.8$\pm$0.1] & [0.016$\pm$0.006] & [60] &  \\
     \noalign{\smallskip}
 \multicolumn{3}{l}{{\bf Detectable binaries}}         & \\
4U1820--30$^{24,25}$    &  UCXB  &   685(4)    &   2.7896  & --7.9144     & - &  3.7$^{45}$ &  [1.4]     &  [0.069]  & [60]  \\
ZTFJ0127$^{26}$            &  DWD$^*$  &  822.680314(43)  & 128.4671  &   --9.5102  &. $0.191$  &  9.26 &  $0.75\pm0.06$  &  $0.19\pm0.03$  & [75-90] \\
SDSSJ2322$^{27}$    &  DWD  & 1201.4(5.9)   &   85.9507   &  --51.2104   & $-0.179$ &  9.08 &  $0.34\pm0.02$  & $>0.17$  &  [60] \\
PTFJ0533$^{28}$   &  DWD  &  1233.97298(17)  &  201.8012	&  --16.2238   & $-0.067$  &  8.70 &   $0.652^{+0.037}_{-0.040}$ & $0.167\pm0.030$ & $72.8^{+0.8}_{-1.4}$ \\
ZTFJ2029$^{9}$    &  DWD$^*$   &   1252.056499(41)   &  58.5836    &  --13.4655  & $-0.054$  &  10.27 &  $0.32\pm0.04$  &  $0.30\pm0.04$ &  $86.64^{+0.70}_{-0.40}$ \\
PTF1J1919$^{29}$   &  AMCVn$^*$    & 1347.354(20)  &   79.5945  &   15.5977   & $0.036$  & 9.08  & [0.8$\pm$0.1] & [0.066$\pm$0.026] & [60] \\
TIC378898110$^{30}$  & AMCVn     &  1347.96     & 297.0555 &  1.9451 & $0.027$ & 6.82 & [0.8$\pm$0.1] & [0.1$\pm$0.02] &  $74\pm10$ \\
CXOGBSJ1751$^{31}$ &  AMCVn    &  1374.0(6)  & 359.9849  &  --1.4108   & $1.623$  &  6.01 & [0.8$\pm$0.1] &  [0.064$\pm$0.026] & [60] \\
ZTFJ0722$^{9}$    &  DWD$^*$  &   1422.548655(71)   &  232.9930   & --1.8604   & $-0.157$  &  8.23 &  $0.38\pm0.04$  &  $0.33\pm0.03$ & $89.66\pm0.22$ \\
%SDSS J010657.39--100003.3  &  135.7244  & --72.4861  & 2345.8   &  0.188$\pm$0.011 &  ${0.57^{+0.22}_{-0.07}}$ & 67$\pm$13 & 18.21 \\
KL Dra$^{32}$  &  AMCVn  &  1501.806(30)  &   91.0140  &   19.1992 & $0.668$  &  9.25 &   0.76 & 0.057 & [60]  \\ 
PTF1J0719$^{33}$ &  AMCVn  &  1606.2(1.2)  &  168.6573  &   24.4945   & $0.425$  & 9.30  &   [0.8$\pm$0.1] & [0.053$\pm$0.021] & [60]   \\ 
CP Eri$^{34,35}$  &  AMCVn  &  1740(84)   &  191.7021 & --52.9098   & $0.218$ &  10.5 &  [0.8$\pm$0.1]  &  [0.049$\pm$0.020]   & [60]  \\
SMSSJ0338$^{21}$ &  DWD  & 1836.1(31.9)    & 128.8576  &  20.7792  & $-0.129$ & 8.64  &    $0.230\pm0.015$    &  $0.38^{+0.05}_{-0.03}$   &  $69\pm9$   \\
J2322+2103$^{20}$ & DWD  & 1918.08(21.60)  &  96.5151 &  --37.1844  & $0.071$ &  8.81 &    $0.291\pm0.013$    &  $0.75\pm0.26$  &  [60]   \\
SDSSJ0106$^{36}$ &  DWD  & 2345.76(1.73)    & 191.9169  &  31.9952  & $-0.223$ &  10.30 &    $0.188\pm0.011$   &  $0.57^{+0.22}_{-0.07}$  &  $67\pm13$  \\
SDSSJ1630$^{37}$   &  DWD  &  2388.0(6.0)  &   67.0760  &   43.3604    & $-0.147$  &  9.54  &  0.298$\pm$0.019 &  0.76$\pm$0.24 & [60]  \\
J1526m2711$^{38}$  &  DWD  &   2417.645(37.930)  &  340.4437  &  24.1935 &  $-0.108$  &  9.36  &   $0.37\pm0.02$  & $>0.40\pm0.02$ & [60]  \\
%J1506m1125$^{35}$  &  DWD  &   2792.448(33.696)  &  347.7034  &  39.4610 &  $-0.195$  &  8.93  &  $0.43\pm0.02$  & $>0.18\pm0.01$ & [60]  \\
SDSSJ1235$^{39,40}$    &  DWD  &  2970.432(4.320)  &   284.5186  &  78.0320    & $-0.219$ & 9.27  &  $0.35\pm0.01$  &  $0.27^{+0.06}_{-0.02}$ & $27.0\pm3.8$ \\
SDSSJ0923$^{41}$  &  DWD   &  3883.68(43.20)   &  195.8199  &   44.7754    & $-0.236$  &  8.62 &   0.275$\pm$0.015 &   0.76$\pm$0.23 & [60]  \\
CD--30$^\circ$11223$^{42}$  &  sdB$^*$   &  4231.791855(155)   &   322.4875  &   28.9379     & $-0.388$  & 4.55  &  0.54$\pm$0.02     &  0.79$\pm$0.01  & 82.9$\pm$0.4  \\
%WD0957   & DWD   &  5269.8108038(725)   &    287.1438   & --9.4629  &  $0.37\pm0.02$  & $0.32\pm0.03$  & $75\pm15$  &  \\
SDSSJ1337$^{43}$ &  DWD  & 5942.952(300)    & 89.0428  &  74.0799  & $0.306$ & 11.31  &  $0.51\pm0.01$  &   $0.32\pm0.01$  & $13\pm1$   \\
HD265435$^{44}$  &  sdB  & 5945.917432(280)  &   87.0170  & 1.1225  & $-0.425$  &  3.76 &   $0.63^{+0.13}_{-0.12}$  &  $1.01\pm0.15$ & $64^{+14}_{-5}$ \\

%ZTFJ1749    &  DWD    &  34.5093    &  17.9025   & 1586.03  &  $0.40^{+0.07}_{-0.05}$ &  $0.28^{+0.05}_{-0.04}$  &  $85.45^{+1.40}_{-1.15}$ \\

%SDSS J082239.54+304857.2   &  191.9169  &   31.9953  & 2416.6   &  0.304$\pm$0.014 &  0.524$\pm$0.050 & {88.1$^{+2.3}_{-1.4}$}  & 23 \\
%SDSS J104336.27+055149.9   &  242.1031  &   52.9139  & 2738.9   &  0.183$\pm$0.010 &  0.76$\pm$0.24 & [60]  & 23 \\
%SDSS J105353.89+520031.0   &  156.4021  &   56.7940  & 3677.2   &  0.204$\pm$0.012 &  0.75$\pm$0.24 & [60] & 18 \\
%SDSS J005648.23--061141.5  &  126.6604  & --69.0278  & 3748.0   &  0.180$\pm$0.010 &  0.82$\pm$0.14 &  [60] & 18 \\
%SDSS J105611.02+653631.5   &  140.0670  &   47.5033  & 3759.3   &  0.334$\pm$0.016 &  0.76$\pm$0.24 & [60] & 18 \\

%SDSS J143633.28+501026.9   &   89.0112  &   59.4607  & 3957.1   &  0.234$\pm$0.013 &  0.78$\pm$0.23 & [60] & 18 \\
%SDSS J082511.90+115236.4   &  212.5705  &   26.1227  & 5027.6   &  0.278$\pm$0.021 &  0.80$\pm$0.22 & [60]  & 18 \\
% WD 0957--666              &  287.1438  &  --9.4629  & 5269.8   &  0.37$\pm$0.02   &  0.32$\pm$0.03 & 50-86  & 24 \\
%SDSS J174140.49+652638.7   &   95.1544  &   31.7085  & 5279.9   &  0.170$\pm$0.010 &  1.17$\pm$0.07 & [60]  & 18 \\ 
%WD 1242--105          & 300.3 & 52.0    & 10260.0   &    0.56$\pm$0.03 &  0.39$\pm$0.02 & 45.1  & 25 \\
%      \noalign{\smallskip}
%  \multicolumn{3}{l}{{\bf Undetectable}}      & \\
%OW J074106.07--294811.0   &   244.8283  &  --3.4737   & 2679.8   &   0.23$\pm$0.12     &  0.72$\pm$0.17  & 57.4$\pm$4.7  & 26 \\
%  \multicolumn{3}{l}{{\bf Ultracompact X-ray binaries}}      & \\
\hline
\end{tabularx}
\begin{flushleft}
\footnotesize{[1]\citet{str05}, [2]\citet{roe10}, [3]\citealt{bur19}, [4]\citealt{bur20a}, [5]\citet{ram02}, [6]\citet{esp05}, [7]\citet{bro11}, [8]\citet{her12}, [9]\citealt{bur20}, [10]\citet{gre18},  [11]\citet{ski99}, [12]\citet{roe06}, [13]\citet{fon11}, [14]\citet{kup15}, [15]\citet{pat02} , [16]\citet{roe07c}, [17]\citet{bro16}, [18]\citet{kil14}, [19]\citet{kos23}, [20]\citet{bro22},  [21]\citet{pro97}, [22]\citet{kil21}, [23]\citet{and05}, [24]\citet{ste87a}, [25]\citet{che20}, [26]\citet{bur23}, [27]\citet{bro20}, [28]\citet{bur19a}, [29]\citet{lev14}, [30]\citet{gre24}, [31]\citet{wev16} [32]\citet{woo02}, [33]\citet{lev11}, [34]\citet{how91}, [35]\citet{gro01}, [36]\citet{kil11}, [37]\citet{kil11a}, [38]\citet{kos23a}, [39]\citet{kil17}, [40]\citet{bre17}, [41]\citet{bro10}, [42]\citet{gei13}, [43]\citet{cha21}, [44]\citet{pel21}, [45] $M_V$ taken from \citet{par94}}
% ,,  [15]\citet{roe07}, , , [20]\citet{kil11a}, [21]\citep{bro10}, [22]\citet{gei13}
% , [16]\citet{cop11}, [17]\citet{nat81},
%, , [21]\citet{kil11}, ,
%[23]\citet{bro17}, [24]\citet{mor97}, [25]\citet{deb15}, [26]\citet{kup17a}, 
%[10]\citet{lev14},[13]\citet{woo02},
\label{tab:system}
\end{flushleft}
\end{center}
\end{table}

\section{Methods} 

\subsection{Improvements from {\it Gaia} DR2 to {\it Gaia} DR3}
In 2018 {\it Gaia} DR2 released full astrometric solutions, including parallaxes, and proper motions for 1.3 billion sources \citep{gai16, gai18}. The release was based on observations taken between July 2014 and May 2016. The parallaxes allowed us for the first time to calculate distances for a large sample of {\it LISA} detectable compact binaries. The distances in combination with the chirp mass provided the opportunity to calculate GW amplitudes and estimate the detectability for {\it LISA} \citep{kup18}. Until {\it Gaia} DR2 only a small sample of AM\,CVn binaries had parallax measurements using the Hubble Space telescope \citep{roe07c}.

About two years after the second data release, the {\it Gaia} early data release (eDR3) provided full astrometric solutions for 1.4 billion sources based on 34\,month of observations \citep{gai20}. The released data included one additional year of {\it Gaia} data leading to a higher precision in the parallaxes and proper motions compared to DR2 as well as first time parallax measurements for several {\it LISA} sources, including ZTFJ1905, ZTFJ0127, SDSSJ0935, V803~Cen and CR~Boo. Generally, we find that the parallaxes improved by about 20\% - 30\% between DR2 and eDR3, in particular for faint sources (Fig.\,\ref{fig:parallaxerrDR2vEDR3}). Our list also contains six binaries with either a negative parallax or a parallax error close to $100\%$. These are ZTF\,J1539, ZTF\,J2243, V407\,Vul, ZTF\,J1905, 4U\,1830-30, and ZTF\,J2029. We anticipate that for these objects the distance estimate is dominated by the derived scale length prior (cf. Section~\ref{sec:distance}). We note that distances can also be estimated through indirect methods. In particular spectroscopic distances have been used for double white dwarfs. For this work we only include parallaxes to derive distances to be independent from spectroscopic models.

In June 2022 {\it Gaia} data release 3 (DR3) was released. {\it Gaia} DR3 included the same data as eDR3 and as such astrometric solutions did not improve between eDR3 and DR3. However, DR3 included a large amount of additional information, including orbital astrometric solutions for wide binaries with a clean solution \citep{gai22}. For the remainder of the paper we will always refer to DR3 knowing that parallaxes and proper motions are the same for eDR3 and DR3.

%Since {\it Gaia} DR2, several additional systems have now a published parallax. This includes ZTFJ1905, SDSSJ0935, V803~Cen and CR~Boo. We find that 21 {\it LISA} binaries from our list with parallax precision better than $20$\%. %For these binaries the inferred distances are independent of the assumed scale length prior. 
%Note that our list also contains six binaries with either a negative parallax or a parallax error close to $100\%$. These are ZTF\,J1539, ZTF\,J2243, V407\,Vul, ZTF\,J1905, 4U\,1830-30, and ZTF\,J2029. We anticipate that for these objects the distance estimate is dominated by the derived scale length prior. %Our distance estimates for ZTF\,J1539, ZTF\,J2243, and ZTF\,J2029 are consistent with the spectroscopic distance \citep{bur20,bur20a} giving further confidence in the assumed prior. 

\subsection{Systems with uncertain parallax or alternative distance estimates}\label{sec:uncertain_parllax_systems}

Here, we take the opportunity to discuss a few systems with uncertain parallax or alternative distance estimates: HM~Cnc, 4U\,1830$-$30, V407\,Vul, and ZTF\,J1905.

HM~Cnc is the only remaining system with no parallax measurement. Therefore, the distance estimate of HM~Cnc remains debated. \citet{roe10} estimated a distance of 5\,kpc based on its properties whereas \citet{rei07} estimated a distance of $\approx$2\,kpc based on the observed flux. Most recently \citet{mun23} presented the discovery of $\ddot{f} = (-5.38\pm2.10)\times10^{-27}$Hz\,s$^{-2}$  in HM\,Cnc. They concluded that HM\,Cnc is close to the period minimum and theoretical MESA\footnote{MESA (Modules for Experiments in Stellar Astrophysics) is an open-source 1D stellar evolution code: \url{https://docs.mesastar.org/en/release-r23.05.1/}} calculations find a mass of $\approx$1\,\Msun\  for the accretor and $\approx$0.17\,\Msun\ for the donor. This result is in strong contradiction to the results presented in \citet{roe10} based on a spectroscopic analysis. \citet{mun23} also discussed different ways to measure the distance to HM~Cnc and found values between 2\,kpc and 11\,kpc. This shows the very large uncertainties of the system properties from EM studies. As a consequence, the expected GW amplitude also remains uncertain. %However, we predict that even at a distance of 10\,kpc, HM~Cnc will be detected within the first three months observations and therefore {\it LISA} will provide an independent measurement of its distance

In practice, 4U\,1830$-$30 also lacks a parallax measurement. However, it is located in the globular cluster NGC\,6624, which allows for an independent distance estimate using color-magnitude diagrams with theoretical isochrones, or by using variable stars that follow known relations between their periods and absolute luminosities like RR\,Lyrae stars. NGC\,6624 has a well-measured distance of $7972\pm277$\,pc \citep{baum21}, which we take as the distance for 4U\,1830$-$30. 

V407\,Vul's optical counterpart is dominated by a component that matches a G-type star with a blue variable \citep{ste06}. It is still unknown whether this is a chance alignment or whether V407\,Vul is a triple system where an ultracompact inner binary is orbited by a G-star companion. Companions in orbits with a multi-year orbital period can present themselves in {\it Gaia} DR3 data, either they are listed in the non-single star tables of {\it Gaia} DR3 ({\bf nss\_two\_body\_orbit}) or they have a non-zero value in the {\bf astrometric\_excess\_noise} keyword in the {\bf gaia\_source} table. The latter is non-zero if the astrometric solution shows additional perturbations to a single-source solution which could be an indication of a astrometric wobble if the G-star in V407\,Vul is in a wide orbit \citep[e.g.][]{bel20,pen20}. V407\,Vul is not listed in the non-single star tables of {\it Gaia} DR3 ({\bf nss\_two\_body\_orbit}) and has an {\bf astrometric\_excess\_noise}=0 and {\bf astrometric\_excess\_noise\_sig}=0 and therefore there is no indication in the current {\it Gaia} DR3 data set that the G-star is a wide companion to the inner ultracompact binary. However, we note that {\it Gaia} DR3 is only sensitive to few years periods. Longer periods would not yet show up as astrometric wobble and therefore we cannot exclude that the G-star has a period of more than a few years.

\begin{figure}
  \begin{center}
    \includegraphics[width=0.48\textwidth]{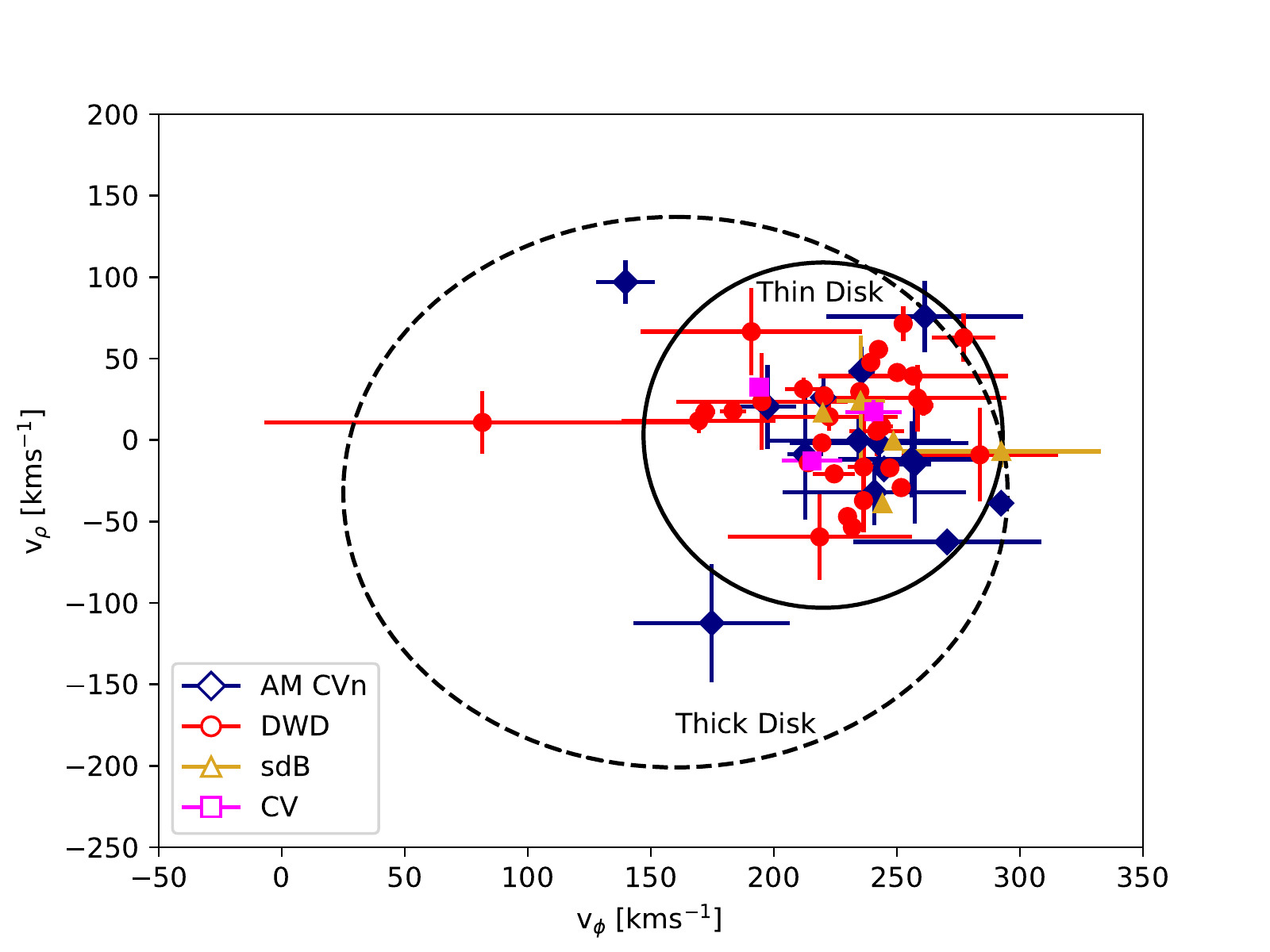}    
    \includegraphics[width=0.48\textwidth]{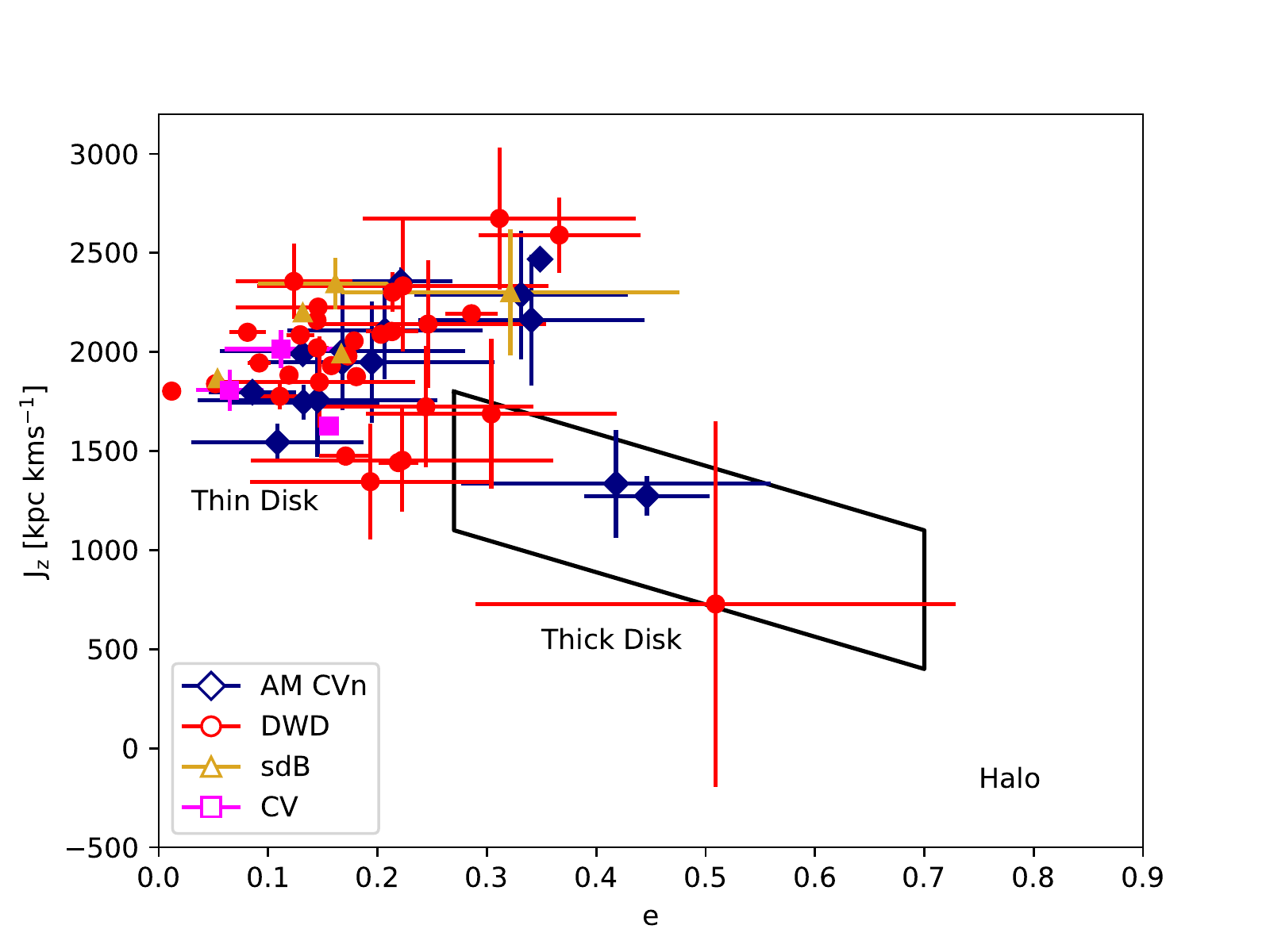}
\vspace{2mm} 
  \caption{$V_\phi$--$V_\rho$ (left) and $e$--$J_{z}$ diagrams (right). The solid and dotted ellipses render the 3$\sigma$ thin and thick disk contours in the $V_\phi$--$V_\rho$ diagram, while the solid box in the $e$--$J_{z}$ marks the thick disk region as specified by \citet{pau06}}
    \label{fig:pop}
    \end{center}
\end{figure}

%Valeriya: I DON KNOW WHERE TO PUT THIS PART
%ZTF\,J1905 has an uncertain parallax in {\it Gaia} DR3 ($\varpi = 1.8652\pm1.5428$) and we find a distance of $697\pm605$\,pc using $L=400$\,pc based on its Galactic kinematics. This results in an absolute magnitude of 11.47\,mag which is inconsistent with AM\,CVn binaries with similar orbital periods such as AM\,CVn or SDSS\,J1908. In the orbital period range between 15\,min - 20\,min, AM\,CVn systems have typically absolute magnitudes around 6.5\,mag. If ZTF\,J1905 has an absolute magnitude of 6.5\,mag it would be located at around $6.5$\,kpc which is an order of magnitude inconsistent with our distance estimate from {\it Gaia} DR3 parallax. There is no indication in the {\it Gaia} DR3 quality keywords that the parallax is problematic. Additionally, there is only modest extinction towards the direction of ZTF\,J1905. \citet{gre19} reports $E(g-r)=0.18$ which results in an extinction of $\approx$0.5\,mag in the Gaia$-G$ band which is not sufficient to explain the large discrepancy. However, if ZTF\,J1905 is located at 6.5\,kpc it would not be a detectable {\it LISA} source.

ZTF\,J1905 presents a particular challenge with its uncertain {\it Gaia} DR3 parallax ($\varpi = 1.8652\pm1.5428$). This parallax value would imply the system's absolute magnitude of $\approx11$\,mag, which seems contradictory when compared to AM\,CVn binaries with similar orbital periods such as AM\,CVn or SDSS\,J1908. Typically, AM\,CVn systems in the orbital period range between 15\,min - 20\,min are much brighter, characterized by absolute magnitudes around 6.5\,mag. There is no indication in the {\it Gaia} DR3 quality keywords that the parallax is problematic. Additionally, there is only modest extinction towards the direction of ZTF\,J1905. \citet{gre19} reports $E(g-r)=0.18$ which results in an extinction of $\approx$0.5\,mag in the Gaia$-G$ band, which is not sufficient to explain the apparent discrepancy. The distance of ZTF\,J1905 thus needs further investigation.

\subsection{Distance estimation} \label{sec:distance}

{\it Gaia} DR3 provides parallaxes which can be used to determine distances. To estimate distances from the measured parallaxes we use a probability-based inference approach \citep[e.g.][]{bai15,igo16,ast16,bai18,luri18,bai21}. We follow a similar approach as described in detail in Sec. 3.2 in \citet{kup18}. The measured parallax follows a probability distribution and with a prior on the true distance distribution for the observed sources we can constrain the distance even if the parallax has large uncertainties. If the parallax uncertainty is below $10$\% - $20$\% the distance estimate is independent of the prior. At larger uncertainties on the distance, the distance estimate becomes more and more dependent on the prior. We apply Bayes' theorem to measure the probability density for the distance:
\begin{equation}  \label{eqn:Bayestheorem}
P(r|\varpi,\sigma_{\varpi})= \frac{1}{Z}\  P(\varpi|d,\sigma_{\varpi})\  P(d),
\end{equation}
where $d$ is the distance, $P(\varpi|r,\sigma_{\varpi})$ is the
likelihood function, that can be assumed Gaussian \citep{lin18}. $P(r)$ is the prior and $Z$ is a normalization constant. As in \citet{kup18} we adopt an exponentially decreasing volume density prior $P(d)$

\begin{equation}  \label{eqn:prior}
P(d) = \begin{cases}\frac{ d^2}{2L^3} \exp(-d/L) & \text {if} \  d > 0  \\
0 & \text{otherwise}, \end{cases}
\end{equation}
where $L > 0$ is the scale length. Compared to our previous study, here we assume two values for $L$ based on the binary's membership to the thin or thick disc as detailed below. We also stress that for systems with poor parallax measurement, the distance estimate largely depends on our assumption for $L$.

To estimate if our candidate {\it LISA} binaries are members of the thin or thick disc, for each binary we calculate Galactic kinematics, i.e. velocity components and Galactic orbit. To do this, sky position, proper motions, systemic velocities, and distances are needed. We extract proper motions from {\it Gaia} DR3 and calculate the distance using Eqs.~\eqref{eqn:Bayestheorem} and \eqref{eqn:prior} by setting $L=400$\,pc or $L=795$\,pc; these are typical values for $L$ for thin disc and thick disc objects respectively\footnote{taken from the {\it Gaia} early data release 3 documentation \url{https://gea.esac.esa.int/archive/documentation/GEDR3/Data_processing/chap_simulated/sec_cu2UM/ssec_cu2starsgal.html}}. Additionally, we use published systemic velocities, typically from radial velocity measurements. For systems with unknown systemic velocities, we assume $1$\,km\,s$^{-1} \pm 50$\,km\,s$^{-1}$. As for the distance estimate, we calculate Galactic kinematics assuming $L$ values of 400\,pc (thin discs) and 795\,pc (thick disc). Using the Galactic potential of \citet{all91} as revised by \citet{irr13}, we calculate velocity in the direction of the Galactic center ($V_\rho$) and the Galactic rotation direction ($V_\phi$), the Galactic orbital eccentricity ($e$), and the angular momentum in the Galactic $z$ direction ($J_z$). The Galactic radial velocity $V_\rho$ is negative towards the Galactic center, while stars that are revolving on retrograde orbits around the Galactic center have negative $V_\phi$. Stars on retrograde orbits have positive $J_z$. Thin disk stars generally have very low eccentricities $e$. Population membership can be derived from the position in the $V_\rho$ - $V_\phi$ diagram and the $J_z$ - $e$ diagram \citep{pau03, pau06}. We find that for all objects the population membership is independent of the assumption for $L$ and we apply the appropriate value for $L$ for the distance estimation. Fig.\,\ref{fig:pop} shows the population memberships for our candidate {\it LISA} sources. Most systems can be identified as thin disc objects. Table\,\ref{tab:system1} presents the calculated distances with the assigned value for $L$.

\begin{table}
%\begin{center}
\caption{Measured EM properties (parallax, distance) and derived GW parameters of the verification binaries and detectable binaries. The distance for HM Cnc is assumed. The fractional error for the amplitude ($\sigma_\mathcal{A}/\mathcal{A}$) and the precision the inclination ($\Delta\iota$) is calculated for four years integration with \emph{LISA}.} 
\begin{tabular}{lrrrrrrr}
\hline
%Source                  &  f$_{\rm GW}$  &  $\pi$ & $\sigma$ & Distance & $\sigma$ & $\mathcal{A}$  & SNR  \\
Source                  &  $f$  &  $\varpi$ DR3$^a$  &  $\varpi$ DR2$^a$  &   $L$  & $d$ & $\sigma_\mathcal{A}/\mathcal{A}$  & $\Delta\iota$  \\
                        &  (mHz)   &  (mas)     & (mas)  &  (pc)  & (pc) &  (\%)   & (deg)  \\
\hline
 \multicolumn{3}{l}{{\bf Verification binaries}}      &     \\
HM Cnc                      & 6.220  &  -    & -    &  & [5000 -- 10,000]  & 9.5 & 21  \\
ZTFJ1539                 &  4.822 &   $-0.4926\pm0.5726$    &  $-0.1125\pm0.7884$      &  795  &  $2469\pm1253$  &  1.1 &  0.6 \\
ZTFJ2243               &  3.788  &    $-1.2372\pm0.6578$   &  $-1.5658\pm1.0522$ &  400  & $1756\pm726$   & $0.36$ & 0.6 \\
V407 Vul                    & 3.512  &  $0.0978\pm0.2384$  & $0.0949\pm0.3272$ & 400  &  $2089\pm684$ & 2.0 & 1.9 \\
ES Cet                   & 3.225  &  $0.5606\pm0.0677$  & $0.5961\pm0.1081$ & 795  &  $1779\pm234$ & 2.2 &  2.1 \\
SDSSJ0651                & 2.614  &  $1.0071\pm0.3091$  &  $1.0002\pm0.4759$  & 400    & $958\pm370$ &  1.8 & 0.9 \\
ZTFJ0538              & 2.308  &    $0.9617\pm0.2866$   &  $1.1477\pm0.4811$     & 400  &  $999\pm366$ &  2.8 & 1.5 \\
SDSSJ1351                & 2.130  &  $0.6584\pm0.2197$  & $0.5957\pm0.3134$ & 795  & $1530\pm755$ &  27.9 & 33 \\
AM\,CVn                     & 1.944  &  $3.3106\pm0.0303$  & $3.3512\pm0.0452$ & 795  &  $302\pm3$ & 12.5 &  23 \\
ZTFJ1905              & 1.938  &  $1.8652\pm1.5428$     &   -    & 400  & $697\pm605$ & 11.6  &  35 \\
SDSSJ1908                   & 1.843  &  $1.0232\pm0.0335$  & $0.9542\pm0.0464$ & 400  &  $977\pm32$ &  19.5 &  30 \\
HP\,Lib                      & 1.814  &  $3.5674\pm0.0313$  & $3.6225\pm0.0525$ & 400  &  $280\pm3$ &  13.3 &  24 \\
SDSSJ0935                   & 1.683  &  $2.7034\pm0.6648$  & -     &  400 & $395\pm203$  &  3.3 & 3.1 \\
J0526+5934                &   1.625  & $1.1826\pm0.0910$   & $1.1294\pm0.1097$ & 400  & $845\pm68$  &  18.0 & 9.6 \\
J1239-2041              &   1.481  &   $1.0068\pm0.2309$    &   $1.4054\pm0.3297$    &  400  &  $972\pm272$  &  13.3 &  9.5 \\
TIC378898110             &  1.483  &   $3.2328\pm0.0195$    &   $3.2186\pm0.0307$    &  400   &  $309\pm2$   &      &   \\
CR Boo                & 1.359  &  $2.8438\pm0.0367$  &  -  & 400  &  $351\pm5$ &  19.8 & 29 \\
SDSSJ0634              &   1.257    &    $2.3111\pm0.0835$   &   $2.3578\pm0.091$    & 400  & $433\pm16$ &  19.4 &  28 \\
V803 Cen                    & 1.253  &  $3.4885\pm0.0599$  &  - & 400  & $287\pm5$ &   16.1 & 27 \\

%ASASSN-14cc              & 22.5 (sh) & &  &  & & & &\\

%KL Dra                      & 1.33  &  1.035  & 0.149  & 956  & 153 & 3.49$\pm$0.53 & 3.7\\
%PTF1J2219+3135           & 26.1 & &  &   & & & & & \\
%PTF1 J071912.13+485834.0    & 1.25  &  1.144  & 0.301  & 861 & 304 & 3.57$\pm$1.91 & 2.8\\
%SDSS J092638.71+362402.4    & 1.18  &  1.824  & 0.549  & 577 & 324 & 3.56$\pm$2.02 & 2\\
%CP Eri                      & 1.17  &  0.684  & 0.941  & 964 & 615 & 2.84$\pm$2.14 & 2\\
%V406 Hya                    & 0.99  &  2.391  & 1.050  & 504 & 493 & 3.96$\pm$3.90 & 2\\
%SDSS J124058.03-015919.2    & 0.89  &  1.857  & 0.612  & 577 & 365 & 2.84$\pm$2.13 & 1\\
%SDSS J012940.05+384210.4    & 0.89  &  2.066  & 0.529  & 507 & 239 & 3.13$\pm$1.94 & 1\\
%SDSS J080449.49+161624.8    & 0.75  &  1.203  & 0.210  & 828 & 173 & 1.36$\pm$0.62 & 0\\
%GP Com                      & 0.72  & 13.731  & 0.060  & 73.0 & 0.4 & 4.56$\pm$1.02 & 2\\
%{\it Gaia} 14aae                  & 0.67  &  3.871  & 0.155  & 259 &  11 & 3.97$\pm$0.27 & 1\\
%SDSS J120841.96+355025.1    & 0.63  &  5.005  & 0.416  & 202 &  18 & 4.06$\pm$1.69 & 1\\
%V396 Hya                    & 0.51  &  10.694 & 0.148  & 94  &   1 & 5.54$\pm$2.25 & 1\\
     \noalign{\smallskip}
 \multicolumn{3}{l}{{\bf Detectable binaries}}         & \\
4U~1820$-$30              &   2.920    &   $-0.7676\pm0.2164$  &  $-0.8199\pm0.2065$ & -  &  $7972\pm277^c$  &   32.3 &  34 \\
ZTFJ0127                &  2.431  &  $ 0.4438\pm0.4657$ &               -     &  400    &  $1283\pm603$  &  7.2  &  2.4 \\
SDSSJ2322              &   1.665    & $1.1558\pm0.2244$  &  $1.2869\pm0.2830$  &  400  & $859\pm206$ &  29.6 &  36 \\
PTFJ0533              &   1.621    & $0.7902\pm0.2396$ & $0.4741\pm0.4786$ &  400  & $1173\pm390$ & 33.3 &  30 \\
ZTFJ2029         &   1.597    & $0.1240\pm0.9893$  & $-1.4269\pm1.4345$ & 400  & $1095\pm644$  & $18.4$  &  10 \\
PTF1J1919        & 1.484  &  $0.6229\pm0.2385$  & $0.5499\pm0.3274$ & 400  & $1364\pm471$ &  65.2 & 46 \\
TIC378898110    & 1.484   & $3.2328\pm0.0195$  &  $3.2186\pm0.0307$ & 400  &  $309\pm2$ &  19.0   &   6.5  \\ 
CXOGBSJ1751       & 1.456  &  $0.8591\pm0.1733$  & $0.4994\pm0.2130$ & 400  &  $1128\pm258$ &  48.4 &  41 \\
ZTFJ0722              &  1.406     & $0.6996\pm0.2457$ & $0.4132\pm0.3508$ & 795  &  $1461\pm785$ & 46.3 & 23 \\
KL\,Dra              &  1.332  &  $1.0817\pm0.0989$ & $1.0354\pm0.1488$ &  795 &  $930\pm91$  &   65.4 &  49 \\
PTF1J0719              &   1.245    & $1.1851\pm0.2292$  &  $1.1436\pm0.3009$  & 400  &  $840\pm201$  &  68.0 &  48 \\
CP\,Eri              &  1.149  &  $1.3451\pm0.2759$ & $0.6836\pm0.9407$ &  400 & $747\pm203$  &  82.6 &   55 \\
SMSSJ0338              &   1.089  & $1.8675\pm0.0562$ & $1.8994\pm0.0793$ & 400  & $536\pm16$ &  43.2 &  40 \\
J2322+2103              &   1.043  &  $0.8261\pm0.2503$ & $0.6943\pm0.3168$ &  400 &  $1134\pm386$ & 38.1 &  38 \\
SDSSJ0106              & 0.853  & $1.2011\pm0.4739$  & $1.3521\pm0.5726$ &  400 & $824\pm441$ & 76.7 &  51 \\
SDSSJ1630        & 0.837  & $1.1748\pm0.1952$  &   $0.9366\pm0.2704$  & 400    & $848\pm167$ &   40.1 & 39 \\
J1526m2711       &  0.827 &  $1.6053\pm0.1751$ &  $1.5683\pm0.2203$   &  400   &  $625\pm75$  & 39.8    &  38 \\
%J1506m1125       &  0.716 &  $2.4219\pm0.1040$  &  $2.3492\pm0.1420$  &  400   & $414\pm18$   &  57.1    &  46 \\
SDSSJ1235              &   0.673  &  $2.2504\pm0.1389$ & $2.3319\pm0.1660$  &  400 & $446\pm28$ &  39.8 & 38 \\
SDSSJ0923     & 0.515  & $3.4795\pm0.0648$  &   $3.3397\pm0.1052$  & 400    & $288\pm5$  &  41.3 & 38 \\
CD--30$^\circ$11223          & 0.473  &  $2.8198\pm0.0516$  &  $2.9629\pm0.0797$   & 400  &  $355\pm7$  &  67.8 & 46 \\
%WD0957              &     &  $6.1130\pm0.0291$  & $6.1121\pm0.0397$ & 400  & $164\pm1$ &   & \\
SDSSJ1337              &    0.337   & $8.8007\pm0.0440$  & $8.6993\pm0.0524$ & 400  &  $114\pm1$  & 32.1 & 35 \\
HD265435              &   0.336    &    $2.1666\pm0.0554$ &  $2.1994\pm0.0628$ & 400  & $461\pm12$  &   70.6 &  54 \\
%SDSS J010657.39-100003.3    & 0.85  & 1.352 & 0.573    & 758  & 460 & 8.29$\pm$5.50 & 3 \\
%SDSS J082239.54+304857.2    & 0.83  & 0.885 & 1.489    & 861  & 628 & 10.33$\pm$7.58 & 3\\
%SDSS J104336.27+055149.9    & 0.73  & -0.054 & 0.393   & 1744 & 684 & 3.92$\pm$1.80 & 1\\
%SDSS J105353.89+520031.0    & 0.54  & 1.506 & 0.445    & 683  & 326 & 9.01$\pm$4.83 & 2\\
%SDSS J005648.23-061141.5    & 0.53  & 1.616 & 0.137    & 620  &  56 & 9.31$\pm$1.51 & 2\\
%SDSS J105611.02+653631.5    & 0.53  & 0.718 & 0.473    & 1104 & 557 & 8.71$\pm$4.85 & 2\\
%SDSS J143633.28+501026.9    & 0.51  & 0.970 & 0.173    & 1011 & 207 & 6.78$\pm$2.06 & 1\\
%SDSS J082511.90+115236.4    & 0.40  & 0.063 & 0.338    & 1786 & 670 & 3.91$\pm$1.69 & 1\\
 %WD\,0957-666               & 0.38  & 6.112 & 0.040    & 163  &  1  & 25.62$\pm$2.33 & 3\\
%SDSS J174140.49+652638.7    & 0.38  & 0.829 & 0.176    & 1159 & 279 & 4.84$\pm$1.22 & 1\\
% WD 1242-105             & 0.19  &   & & & & & \\
\hline
\end{tabular}
\begin{flushleft}
$^a$\citet{gai18}, $^b$\citet{gai20}, $^c$\citet{baum21}
\label{tab:system1}
\end{flushleft}
%\end{center}
\end{table}

\subsection{Gravitational wave parameter estimation} \label{sec:ldsoft}

Gravitational radiation for a typical stellar remnant binary at mHz frequencies 
 can be modeled as a quasi-monochromatic signal characterized by 8 parameters: GW frequency $f$, heliocentric amplitude ${\cal A}$, frequency derivative $\dot{f}$, sky coordinates $(\lambda, \beta)$, inclination angle $\iota$, polarization angle $\psi$, and initial phase $\phi_0$. The GW frequency and amplitude are given by
\begin{equation}
    f = 2/P_{\rm orb},
\end{equation}
with $P_{\rm orb}$ being the binary's orbital period, and
\begin{equation}\label{eqn:amp}
    \mathcal{A} = \frac{2 (G \mathcal{M})^{5/3} }{c^4 d} (\pi f)^{2/3}.
\end{equation}
This is set by the binary's distance $d$ and chirp mass
\begin{equation}\label{eqn:mchirp}
    \mathcal{M} = \frac{(m_1 m_2)^{3/5}}{(m_1 + m_2)^{1/5}}, 
\end{equation}
for component masses $m_1$ and $m_2$.
The frequency derivative $\dot{f}$ is expected to follow the gravitational radiation equation:
\begin{equation}\label{eqn:fdot}
    \dot{f} = \frac{96}{5}\frac{(G\mathcal{M})^{5/3}}{\pi c^5} (\pi f)^{11/3}.
\end{equation}

To forecast {\it LISA} observations of the known binaries we use the {\sc vbmcmc} sampler in {\sc ldasoft}~\citep{PhysRevD.101.123021}.  The sampler uses a parallel tempered Markov Chain Monte Carlo algorithm with delta-function priors on the orbital period and sky location of the binary based on the EM observations (cf. Table~\ref{tab:system}). The priors on the remaining parameters are uniform in log amplitude and cosine inclination with the start value for the inclination being set to the measured values for systems with constraints on the inclination and set to 60\,deg for unconstrained systems. The sampler is also marginalizing over the first time derivative of the frequency, polarization angle, and initial phase of the binary, all of which are considered nuisance parameters for this study. Because this is a targeted analysis of known binaries the trans-dimensional sampling capabilities in {\sc ldasoft} are disabled and the algorithm uses a single template to recover the signal -- effectively a delta function prior on the model. In this configuration, results for binaries below the detection threshold are used to set upper limits on the GW amplitude parameter.

The data being analyzed are simulated internally by {\sc vbmcmc} and include stationary Gaussian noise with the same instrument noise spectrum as was used for the {\it LISA} Data Challenges (LDCs) in Challenge 2a\footnote{\url{https://lisa-ldc.lal.in2p3.fr/challenge2a}} plus an estimated astrophysical foreground from the unresolved Galactic binaries as described in~\cite{Cornish_2017}.  The analysis ignores any correlations, contamination, or additional statistical uncertainty caused by the presence of other signals in the data, and also treats the astrophysical foreground as a stationary noise source. Relaxing these simplifying assumptions will be most relevant for binaries where the astrophysical foreground dominates the instrument noise spectrum at GW frequencies ${\lesssim}3$ mHz but is currently beyond the scope of this analysis (considering overlap with other sources) or capabilities of the sampling algorithm (considering non-stationary noise).

%%%%%%%%%%%%%%%%%%%%%%%%%%%%%%%%%%%%%%%%%%%%%%%%
%%%%%%%%%%%%%%%%%%%%%%%%%%%%%%%%%%%%%%%%%%%%%%%%

\section{Results}

%\begin{figure*}
%  \begin{center}
%    \includegraphics[width=0.8\textwidth]{VBexamples.png}
%  \caption{Three examples of posteriors for binary's inclination and GW amplitude; from top to bottom these are: ZTFJ1539 (verification binary), ZTFJ0538 (detectable binary) and SDSSJ0822 (non-detectable binary). On the left we show how the estimated fractional error on the amplitude $\sigma_{\cal A}/{\cal A}$ and inclination $\sigma_\iota/\iota$ change as the functions of observation time; for comparison we also show current EM constrains and assuming that these won't change before {\it LISA}'s launch (cf. Table~\ref{tab:system} and \ref{tab:system1}).}
%    \label{fig:posterior_examples}
%    \end{center}
%\end{figure*}

\begin{figure*}
 \begin{center}
    \includegraphics[width=0.77\textwidth]{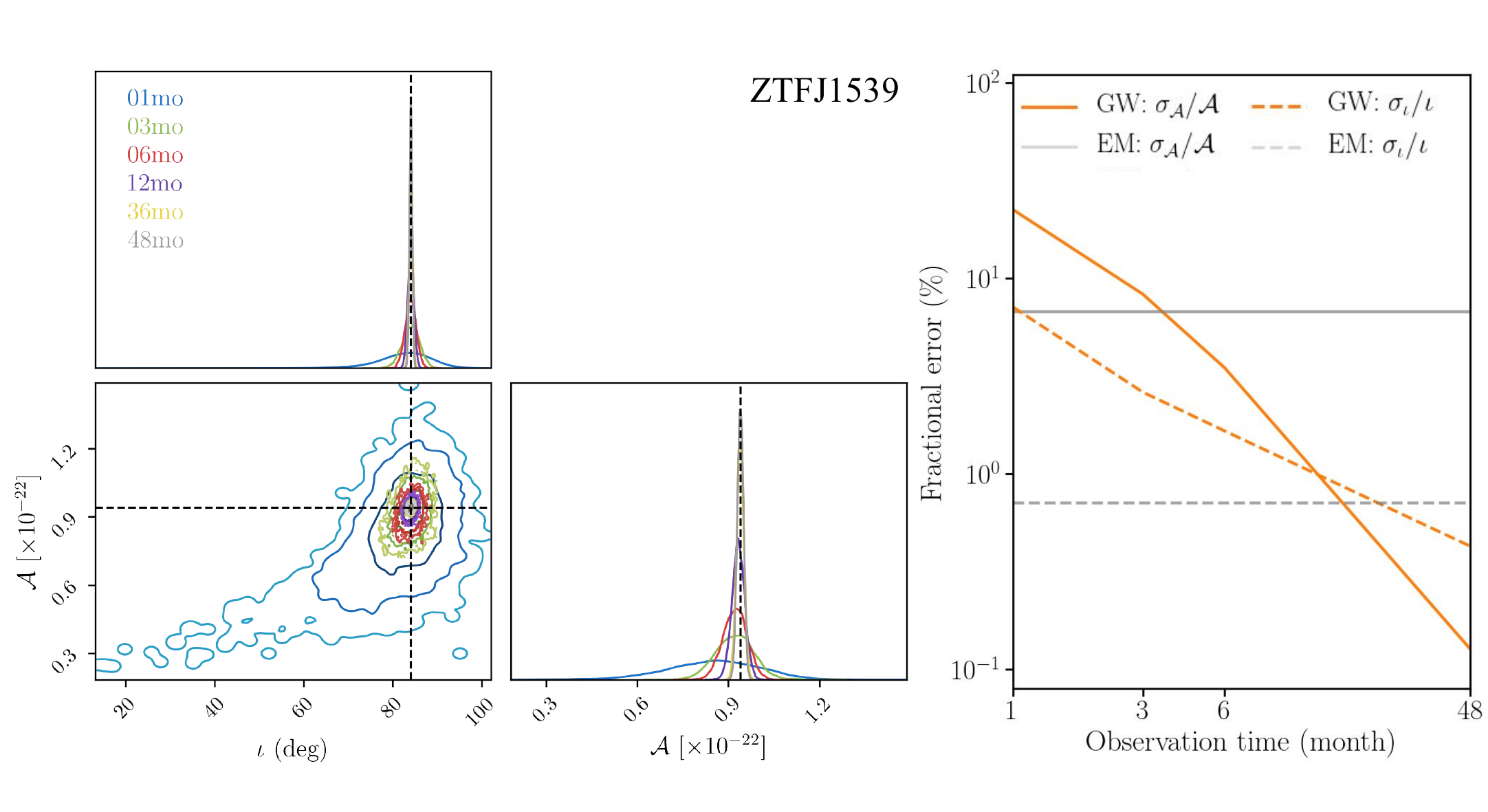}
    \includegraphics[width=0.77\textwidth]{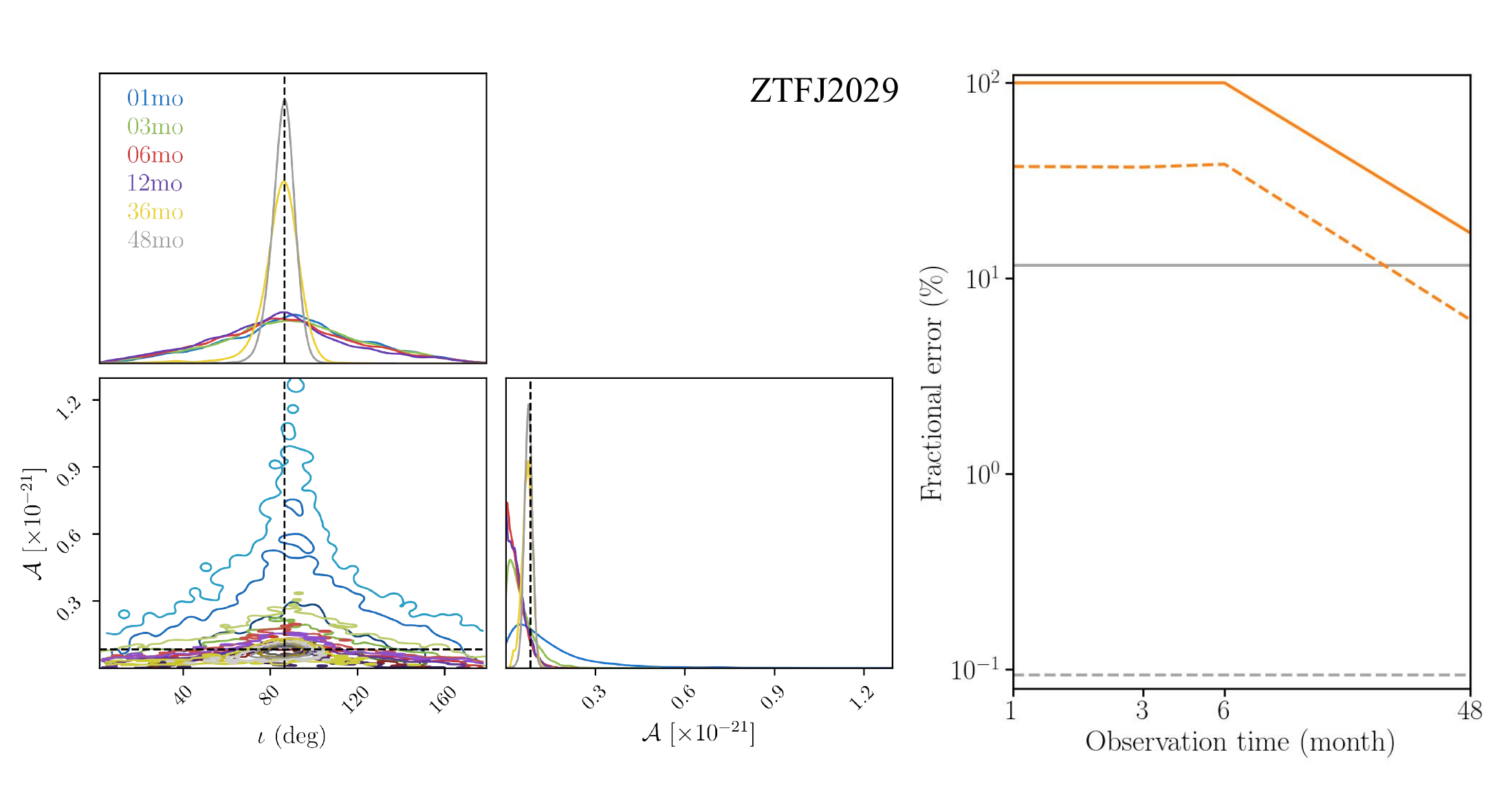}
    \includegraphics[width=0.77\textwidth]{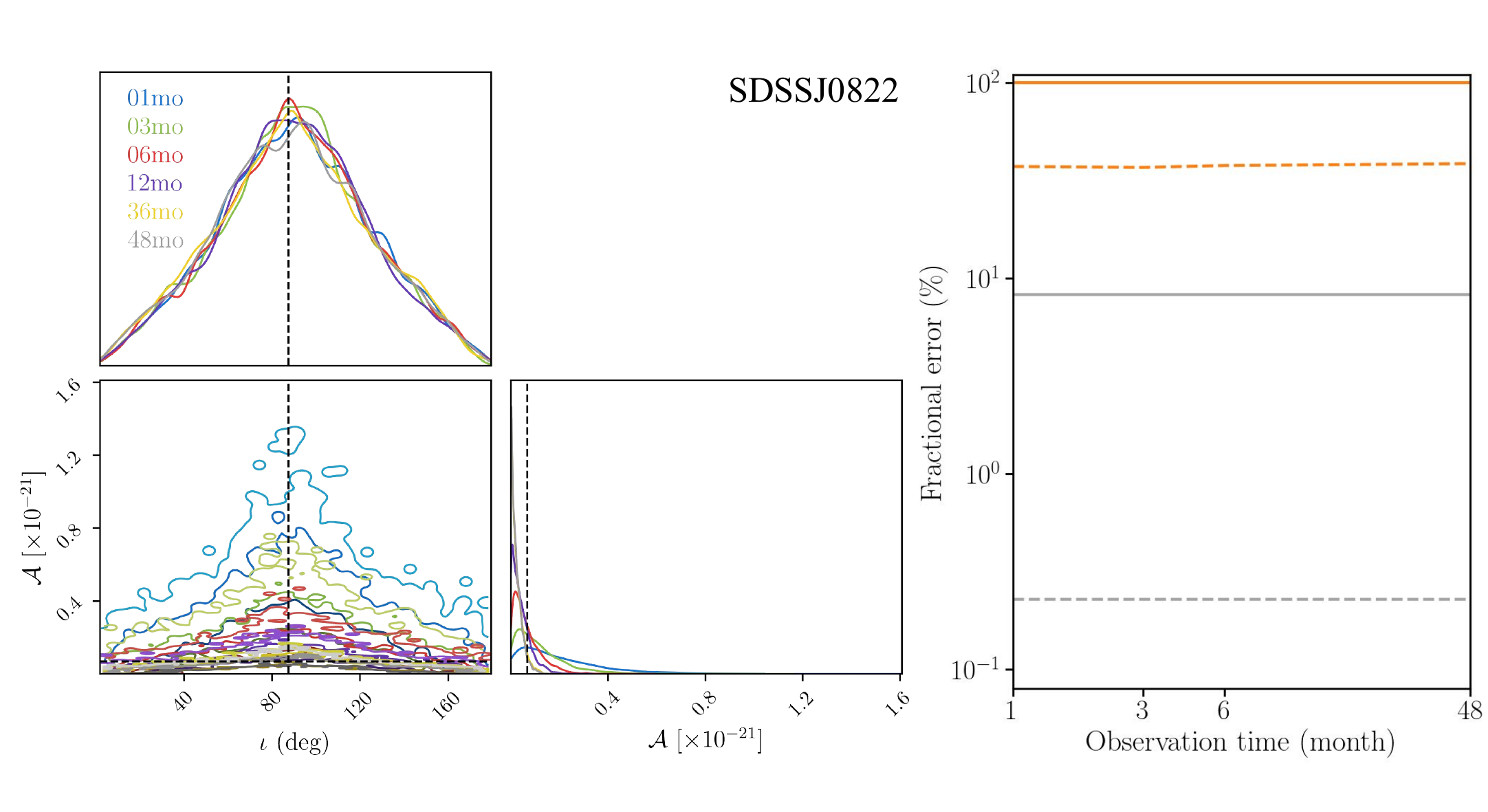}
  \caption{Three examples of posteriors for binary's inclination and GW amplitude; from top to bottom these are: ZTFJ1539 (verification binary), ZTFJ2029 (detectable binary) and SDSSJ0822 (non-detectable binary). On the left we show how the estimated fractional error on the amplitude $\sigma_{\cal A}/{\cal A}$ and inclination $\sigma_\iota/\iota$ change as the functions of observation time; for comparison we also show current EM constrains and assuming that these won't change before {\it LISA}'s launch (cf. Table~\ref{tab:system} and \ref{tab:system1}).}
    \label{fig:posterior_examples}
    \end{center}
\end{figure*}

\begin{figure*}
  \begin{center}
  \includegraphics[width=0.93\textwidth]{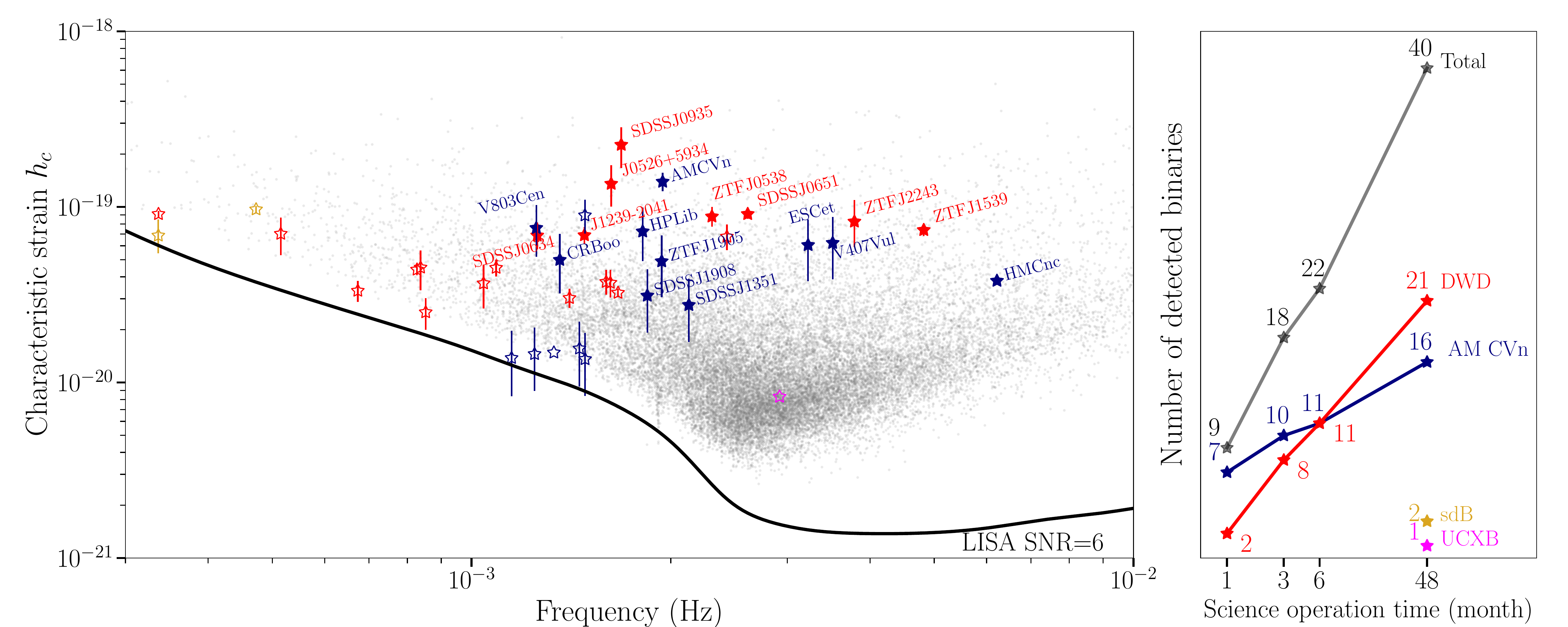}    
\vspace{2mm}
  \caption{{\it Left panel:} Characteristic strain - frequency plot for detectable and verification binaries: AM~CVns in blue, DWDs in red, sdBs in yellow and UCXB in magenta. Filled stars represent binaries detectable within 3 month of observations, which we call here `verification binaries'. The error bars on characteristic strain show $1\sigma$ uncertainty evaluated by generating random samples from EM measurement uncertainties on binary component masses and distances presented in Table~\ref{tab:system} and \ref{tab:system1}.  Black solid line represents {\it LISA}'s sensitivity curve that accounts for the instrumental noise \citep{LISAdoc} and Galactic confusion foreground \citep{Babak:2017tow}. For comparison in gray we show a mock Galactic DWD population detectable with {\it LISA} from \citet{wil21}. {\it Right panel:} Number of detected binaries as a function of science operation time. }
    \label{strainDR3}
    \end{center}
\end{figure*}

We analyze \textbf{55} verification binary candidates in total using the {\sc vbmcmc} sampler in {\sc ldasoft} for increasingly longer {\it LISA} mission science operation time: 1, 3, 6, 12, 36 and 48 months. We remind the reader that  we perform a ``targeted'' analysis by fixing --  using delta-function priors -- binary's sky position and orbital period to the values provided by EM measurements; additionally, we marginalize over $\dot{f}$, $\psi$ and $\phi_0$ (cf. Section~\ref{sec:ldsoft}). Differently from our previous work to assess the detectability of a binary, instead of evaluating the SNR we look at the binary's GW parameters posterior samples. We consider the binary as {\it detectable} if the GW amplitude parameter is constrained away from the minimum value allowed by the prior.  This is most readily identifiable by looking at the two dimensional posterior distribution in the GW amplitude-inclination plane, where a detectable binary will have closed contours in the posterior. We call as {\it verification binary} a system that becomes detectable (as explained in Sec.\,\ref{sec:intro}) within 3 months of observation time with {\it LISA} and we call as {\it detectable binary} when it is detected after 48 months of observation time with {\it LISA}. We introduce this distinction to highlight the use of verification binaries for the early data validations, e.g. in preparation for the first data release. 

Our results show that an integration time as short as 1-3 months would allow for basic consistency tests on the recovered parameters on a few epochs of commissioning data for 9-18 verification binaries; while extending our definition up to 6 month increases the sample by only four additional binaries. To clarify the difference between a verification, detectable and non-detectable binary, in Fig.~\ref{fig:posterior_examples} we show posteriors of the binary's inclination and GW amplitude for three examples: ZTFJ1539 ($\sim7\,$min orbital period, verification binary), ZTFJ2029 ($\sim21\,$min, detectable binary) and SDSSJ0822 ($\sim40$\,min, non-detectable binary). ZTFJ1539 shows closed contours already after 1 month of observations, which become increasingly narrow as observation time increases. Being detectable so early on, it is highly likely that ZTFJ1539 can be used as one of {\it LISA's} verification binary. ZTFJ2029 represents an intermediate case as initially its posterior contours are open at low amplitude, but they close after 36 months at which point we classify this binary as detectable. Finally, we show the case of SDSSJ0822 which, based on the same reasoning as above, we classify as non detectable. %for the posterior distribution for all {\it LISA} sources). 
Right panels of Fig.~\ref{fig:posterior_examples} illustrate how {\it LISA}'s fractional error on GW amplitude ${\cal A}$ and inclination $\iota$ (orange lines) improve over time. Assuming that EM measurement would not improve in the future, which is plausible if no additional EM measurements are taken between now and when {\it LISA} will fly, we show the estimate of the same parameters based on the current EM measurement (gray lines) for comparison.

Based on our definition above, overall we find 40 binaries detectable with {\it LISA} within 4\,yr of science operations, out of which we classify 18 (10 AM~CVns + 8 DWDs) as verification binaries, i.e. detectable within the first 3 months. We list their properties in Table~\ref{tab:system} and \ref{tab:system1}. The summary of our results is presented in Fig.~\ref{strainDR3}.
In the left panel we show characteristic strain - frequency plot for all detectable binaries: AM~CVns in blue, DWDs in red, sdBs in yellow and UCXB in magenta. Filled stars represent {\it LISA} verification binaries, empty stars are detectable binaries within the nominal mission life time (48 month). We compute the error bars on characteristic strain by generating random samples from EM measurement uncertainties on binary component masses and distance (cf. Table~\ref{tab:system} and \ref{tab:system1}); in the figure we plot $1\sigma$ uncertainty based on our random samples. For comparison we also show a mock Galactic DWD population of \citet{wil21} in gray, as well as the {\it LISA} sensitivity curve at SNR=6 as black solid line. The comparison reveals that the current sample of known binaries is mainly representative of the `loudest' GW sources in the Milky Way, while the majority is yet to be discovered in more remote parts of our Galaxy inaccessible for EM observatories \citep[see figure 15 of][]{LISAastroWP}. In the right panel of Fig.~\ref{strainDR3} we show the detection statistic as a function of the science operation time with a break down for different types of binaries showing that in increased science operation for {\it LISA} will lead to a larger number of detected binaries in the {\it LISA} data. We note that in our study none of the CVs will be detectable within four years of {\it LISA} observations.

\begin{figure*}
  \begin{center}
  \includegraphics[width=0.97\textwidth]{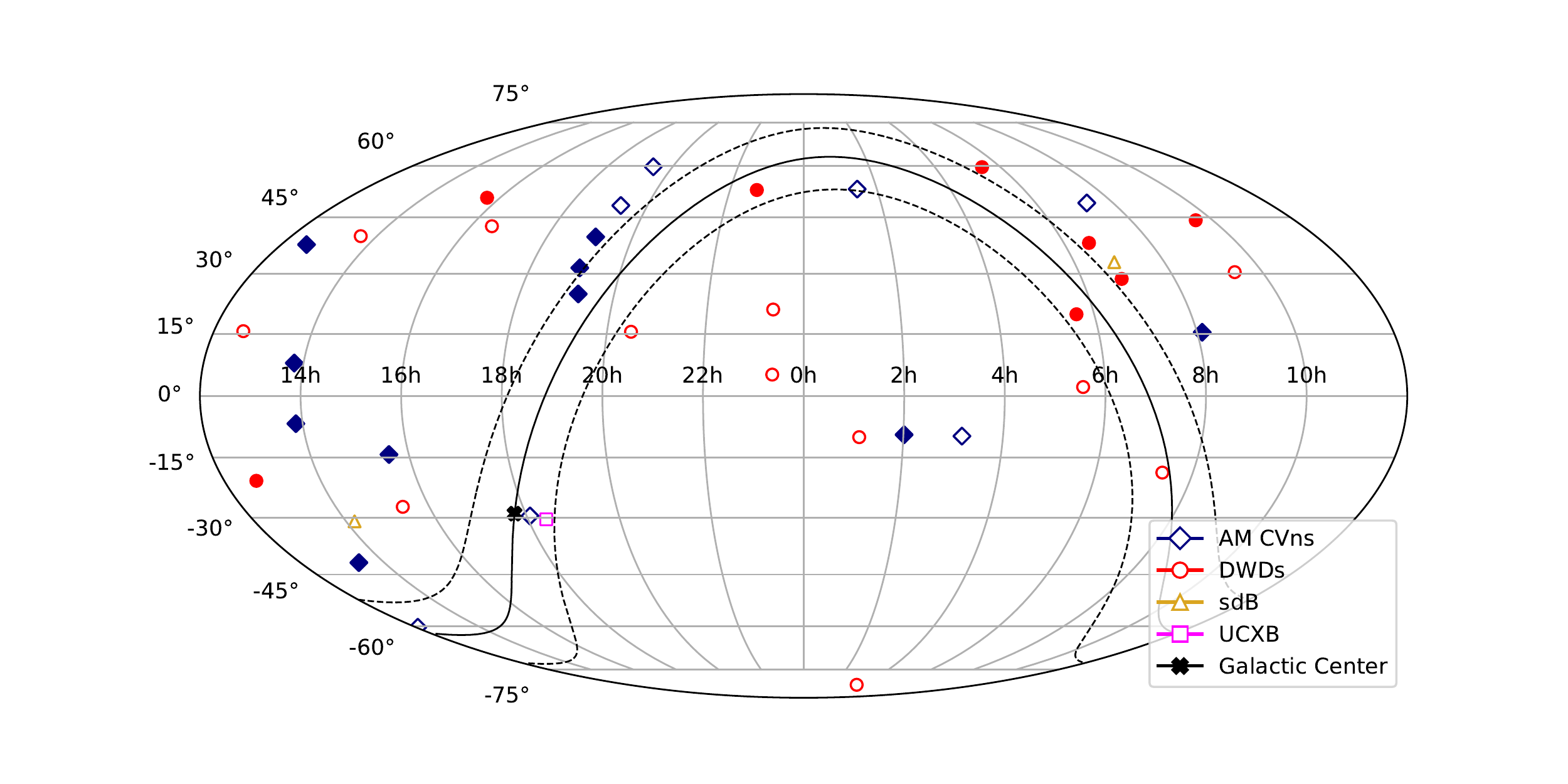}    
\vspace{2mm}
  \caption{The position of the {\it LISA} binaries on the sky in an equatorial projection, with the Galactic plane ($\pm10^\circ$) shown by the full and dashed lines. Filled symbols are verification binaries whereas open symbols are detectable binaries.}
    \label{fig:skymap}
    \end{center}
\end{figure*}

In our previous study \citet{kup18}, we identified 13 detectable binaries. The reasons for the difference are multiple. Firstly, the sample of candidate verification binaries has tripled in the past few years (cf. Section~\ref{sec:sample}). Secondly, some distance estimates have improved between {\it Gaia} DR2 and DR3 (this is, for example, true for CXOGBSJ1751 and SDSSJ163); for some binaries (ZTFJ0127, ZTFJ1905, SDSSJ0935, V803~Cen, CR~Boo) parallaxes were not available as part of the DR2. Most importantly, we also change our criterion for the detectability moving away from a SNR based definition. Recently, \citet{fin23} have used our catalog for a number of {\it LISA} data analysis investigations by performing a Bayesian parameter estimation with the {\sc Balrog} code \citep{roe20,bus21,kle22}. They verified that binaries with SNR$<$6 generally display broad posteriors, with no clear peaks and with amplitude parameter being inconsistent with zero. Thus, they adopted the SNR threshold of 6 as criterion for the detectability with {\it LISA}. They found that up to 14 binaries can be detected within 3 months of observations (see their figures~4 and B1). This result is in agreement with ours considering new binaries that have been added to the list of candidates in our study.%resulting detectable within 3 month (SDSSJ1351, SDSSJ1908 and CR\,Boo) in our study can be explained when taking into account the differences in the data analysis.

We report estimated fractional error on the GW amplitude ($\sigma_{\cal A}/{\cal A}$) and the estimated precision for the inclination ($\Delta \iota$) that can be reached after 4\,years of observation in Table~\ref{tab:system1}.  On average, for verification binaries the amplitude is forecasted to be measured to $\sim10$\%, while the inclination is expected to be determined with $\sim15^\circ$ precision. For detectable binaries these average errors decrease to $\sim50$\% and to $\sim40^\circ$ respectively. From the table one can deduce that the measurements depend on the binary's frequency (generally improving with increasing frequency) and strength of the signal (with verification binaries being better characterized than detectable binaries). Our estimates are in good agreement with \citet[][see their table~2]{fin23}.

\begin{figure}
  \begin{center}
  \includegraphics[width=0.7\textwidth]{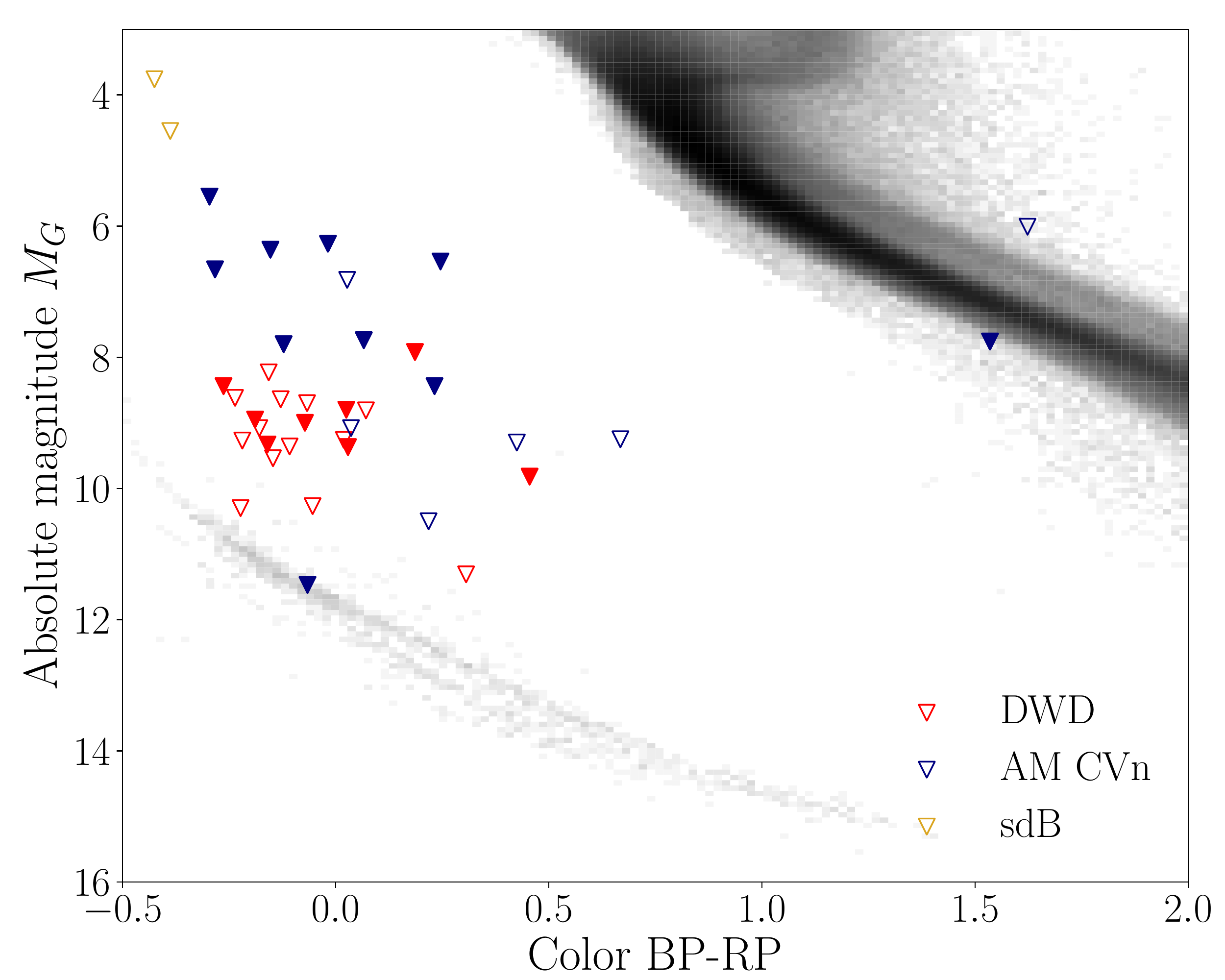}    
\vspace{2mm}
  \caption{%Positions of verification binaries on the {\it Gaia} Hertzsprung-Russell diagram show a clear bias in the current sample towards the brightest and bluest binaries. 
  Verification binaries (in color) on the {\it Gaia} Hertzsprung-Russell diagram (in gray).
  Down-pointing triangles are used to symbolize that absolute magnitude estimates are to be interpreted as upper limits because we do not account for extinction. As before, filled symbols represent verification binaries whereas open symbols represent detectable binaries.} 
    \label{fig:gaiaHRD}
    \end{center}
\end{figure}

Figures~\ref{fig:skymap} and \ref{fig:gaiaHRD} illustrate respectively the position of detectable (empty symbols) and verification (filled symbols) binaries on the sky and on the {\it Gaia} Hertzsprung-Russell (HR) diagram. Both reveal different limitations of the current sample. Figure~\ref{fig:skymap} shows that although the size of the candidate {\it LISA} binaries is progressively growing, it is still biased towards the Northern Hemisphere (where the majority of surveys have been conducted so far), and to high latitudes (to avoid the dust extinction and crowding in the Galactic plane). From observations of bright non-degenerate stars we know that the Galactic stellar population is concentrated in the disc (region in between dashed lines) peaking towards the Galactic center (thick black cross in Fig.~\ref{fig:skymap}), and so we expect {\it LISA} detectable binaries to follow the same distribution \citep[e.g. see][]{sze23}. Figure~\ref{fig:gaiaHRD} shows our sample of {\it LISA} sources compared to the absolute magnitudes ($M_G$) and colors ($BP-RP$) of {\it Gaia}'s sample of stars with parallax uncertainty below 1\% (gray points). The plot shows a bias towards sources brighter and bluer compared to objects on the white dwarf track where most DWDs are expected. %of {\it LISA} detectable binaries. These have been evaluated using (single) WDs with hydrogen-atmosphere cooling models\footnote{White dwarf cooling models are freely available at \url{https://www.astro.umontreal.ca/~bergeron/CoolingModels/}}; we show some of these cooling tracks for a reference. Note however that AM~CVns cannot be directly compared to WD cooling models as for these binaries there is extra luminosity coming from accretion \citep{roe07b,car12,car14a}.
We note that on the HR diagram we use down-pointing triangles to highlight that our absolute magnitude estimates in our sample are to be interpreted as upper limits. This is because for all candidate {\it LISA} binaries the extinction, which is necessary to convert apparent $G$ magnitudes measured by {\it Gaia} magnitudes into absolute magnitudes $M_{G}$, is not well measured. Potential {\it LISA} sources with larger magnitudes can only be observed up to a couple kpc with current surveys, which limits the volume where {\it LISA} binaries can be detected. In contrast to EM searches, {\it LISA} measures directly the amplitude of GW waves, rather than the energy flux. Thus, the observed GW signal scales as $1/d$, rather than $1/d^2$, allowing {\it LISA} to detect binaries at larger distances -- potentially within the entire Milky Way volume -- than in the traditional EM observational bands. 

%%%%%%%%%%%%%%%%%%%%%%%%%%%%%%%%%%%%%%%%%%%%%%%%%%
%%%%%%%%%%%%%%%%%%%%%%%%%%%%%%%%%%%%%%%%%%%%%%%%%%
\section{Discussion}

%In this section we discuss the current sample of {\it LISA} verification and detectable binaries focusing on some systems more in detail. We also provide a future outlook highlighting on what measurements EM studies should focus more in order to maximize the scientific output for this sample. 
\subsection{Limitations of the current sample and prospects to expand the sample}

The currently known sample of candidate {\it LISA} verification binaries is inhomogeneous and biased. This bias is evident in the sky distribution (Fig.~\ref{fig:skymap}), predominantly favoring the Northern hemisphere with $60$\% of {\it LISA} sources. Furthermore, the HR diagram (Fig.~\ref{fig:gaiaHRD}) reveals an over-representation of sources above the main white dwarf track. This is due to the nature of magnitude-limited ground-based surveys favoring brighter sources. Finally, a significant challenge lies in the multitude of detection and analysis methods, leading to a non-uniformity in presenting parameters of these binaries. Parameters are presented in varied ways: with a $1-\sigma$ error, no constraints, approximations or limits. To facilitate more effective future multi-messenger studies, it is essential to develop uniform analysis methods that provide a standard set of prior information.

%However, the commencement of surveys such as ELM Survey South, SDSS-V, BlackGEM, LSST, and {\it Gaia} DR4, which cover the Southern hemisphere and the entire sky, are expected to improve this bias (see Sec.~\ref{sec:future}).

%Future surveys like LSST, Roman, and possibly Euclid, with larger limiting magnitudes, are projected to detect fainter sources on the white dwarf cooling track by probing larger volumes.

%\subsection{Prospects for expanding the sample} \label{sec:future}

Between {\it Gaia} DR2 and {\it Gaia} DR3 the number of detectable {\it LISA} sources has tripled which can be explained by the large number of sky surveys that came online over the last few years. This will improve even more over the next decade with many additional surveys coming online over the next few years. As proven in the past, photometric and spectroscopic surveys are ideal tools to find new {\it LISA} binaries and complement each other. Eclipses or tidal deformation lead to photometric variability on the orbital period, whereas compact {\it LISA} binaries show up in multi-epoch spectroscopy due to large radial velocity shifts between individual spectra.

SDSS-V is an all-sky, multi-epoch spectroscopic survey which started operations in 2020 and will provide spectra for a few million sources \citep{kol17}. Already in early SDSS-V data \citet{cha21} discovered a new detectable {\it LISA} binary. Other spectroscopic ongoing or upcoming spectroscopic surveys include LAMOST \citep{lam12}, 4MOST \citep{jong19} or WEAVE \citep{dal14}. The Asteroid Terrestrial-impact Last Alert System \citep[ATLAS,][]{ton18,hei18} and the Gravitational-wave Optical Transient Observer \citep[GOTO,][]{ste22} are ongoing photometric sky surveys with telescopes located in both hemispheres allowing for an all-sky coverage for both surveys. Their cadence and sky coverage are well suited to find {\it LISA} detectable binaries. BlackGEM is a photometric sky survey covering the Southern hemisphere which has started operations in 2023 \citep{bloem15}. Part of the BlackGEM operations will be the BlackGEM Fast Synoptic Survey which is a continuous high cadence survey of individual fields in the Southern hemisphere. The cadence is ideal to discover new {\it LISA} detectable binaries. First light for the Vera Rubin telescope is expected in 2024. The telescope will perform the Legacy Survey of Space and Time (LSST) covering the Southern hemisphere down to 24\,mag (per single exposure) going much deeper and hence covering a significantly larger volume than current sky surveys. Although the cadence is expected to be not ideal for {\it LISA} binaries, LSST will collect sufficient photometric observations over 10 years to be able to discover {\it LISA} binaries. The next large {\it Gaia} data release (DR4) is expected to include precision time series photometry of $\approx$70 epochs for each object taken over a five year time frame. Euclid is a space mission operating in the near-infrared and visible bands which was launched in 2023 with an unprecedented sky resolution of almost one order of magnitude better compared to ground based observatories. Part of Euclid's operation will be the Euclid deep survey covering a total of 40 sqd for three distinct fields. Each field will get several tens of epochs over its nominal mission time of six years with a depth of $\approx25$\,mag for each epoch in the near-infrared bands \citep{euc11}. Finally, in the late 2020s the Nancy Roman space telescope will conduct a wide-field survey with the same sky resolution as Euclid down to 24 mag. Therefore, we expect that the number of {\it LISA} detectable sources will significantly increase before the {\it LISA} launches providing a large sample of multi-messenger sources \citep[e.g.][]{kor17,li20}.

\begin{figure}[!t]
  \begin{center}
    \includegraphics[width=0.6\textwidth]{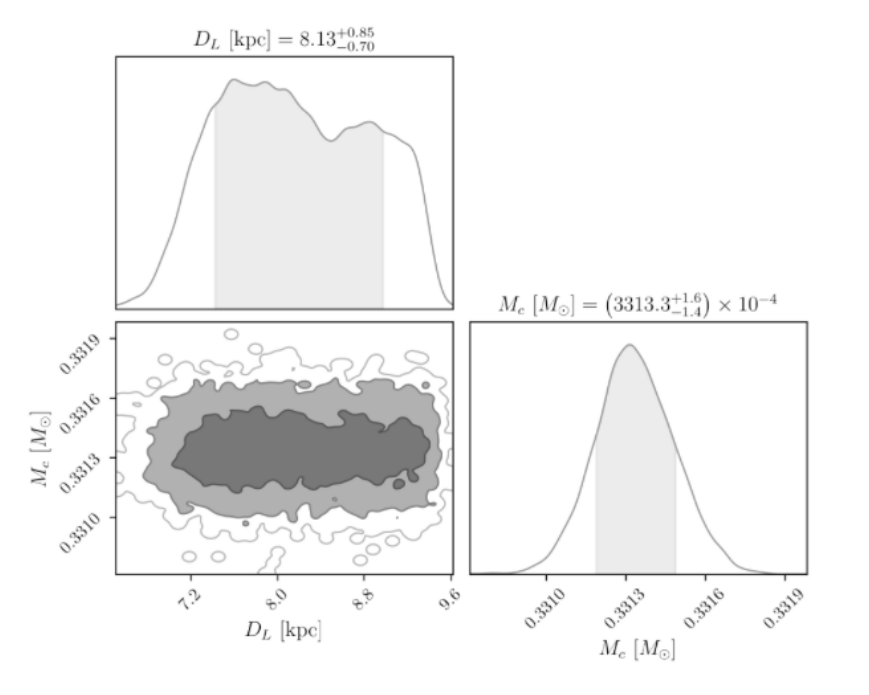}
  \caption{Posterior distribution for HM Cnc's expected chirp mass and distance precision predicted for four years of {\it LISA} observations.}
    \label{fig:hmcnc_dist}
    \end{center}
\end{figure}

%%%%%%%%%%%%%%%%%%%%%%%%%%%%%%%%%%%%% NEW 5.2
\subsection{Focus for future EM efforts}

EM measurement of a binary's inclination will significantly inform {\it LISA}'s data analysis and parameter estimation, particularly for nearly edge-on systems (Table~\ref{tab:system}). As \citet{sha12} outlined, EM inclination data enhances GW amplitude measurement due to the strong correlation between amplitude and inclination parameters in GW data (Fig.~\ref{fig:posterior_examples}). This improves chirp mass and distance determination. Also, \citet{fin23} demonstrated the advantage of EM prior information in reducing detection time versus a blind search. They show that after the binary has been detected, further parameter estimation improvements are inclination-dependent; face-on sources benefit greatly from prior inclination knowledge, with amplitude measurement improvement increasing with observation time (their figure~5). This underscores the importance of binary inclination data from EM observations for future {\it LISA} data analysis. However, 40\% of current sources lack inclination measurement. These are primarily non-eclipsing sources where measurement is non-trivial.

Determining $\dot{f}$ from GW data alone can be challenging for low-frequency $(f<2$ mHz) and/or low-SNR binaries. On the other hand, high-precision $\dot{f}$ measurement is achievable via EM observations for eclipsing sources through ground-based eclipse timing measurements over long baselines (e.g., $>$10 years). \citet{sha14a} demonstrated the accuracy improvement on the chirp mass when adding EM $\dot{f}$ data to GW analysis.

Knowledge of distance ($d$) or parallax ($\varpi$) is also crucial for characterizing {\it LISA} binaries. As mentioned in Section~\ref{sec:distance}, parallax constrains the luminosity distance, directly affecting GW amplitude (Eq.~\ref{eqn:amp}). {\it LISA}'s GW frequency and amplitude measurements, coupled with {\it Gaia}-based parallax, allow binary chirp mass determination (solving Eq.~\ref{eqn:amp} for ${\cal M}$) without measuring $\dot{f}$, usually required for chirp mass determination from GW data (Eq.~\ref{eqn:fdot}). Error propagation illustrates that chirp mass error and parallax error are linearly related ($\sigma_{\cal M}/{\cal M} \propto \varpi \sigma_{\varpi}$), meaning that parallax measurement improvement directly enhances chirp mass estimation. This method recovers chirp masses for binaries at {\it LISA}'s low-frequency end and for interacting binaries, whose $\dot{f}$ contains an astrophysical contribution \citep[e.g.][]{bre18,lit19}, as in verification binaries AM~CVn and HP~Lib. %For systems without a {\it Gaia}  \vk{If we add HM Cnc to distance to the appendix, here we can mention `` If the distance is unknown from EM, like will provide us with distances (see appendix)"}

The forthcoming {\it Gaia} DR4, based on 66 months of data collection, should mark significant improvement over the DR3 (34 months). An estimated improvement factor for parallax errors of $\sim 0.7$ can be expected (using $\sigma_{\varpi} \propto (T_{\rm DR3}/T_{\rm DR4})^{0.5} \sim (34/66)^{0.5}$), with $T_{\rm DR3}$ and $T_{\rm DR4}$ expressed in months, see \citealt{gai18}. Additional enhancements in {\it Gaia} data quality are anticipated, given the indicative mission extension until 2025\footnote{https://www.cosmos.esa.int/web/gaia/release}, allowing for data collection up to a total of 10 years. However, very distant or faint systems might never have a reliable parallax measurement from Gaia. If the period is sufficiently short, {\it LISA} will provide an $\dot{f}$ measurement which can break the degeneracy between chirp mass and distance and as such {\it LISA} will provide an independent distance measurement. Fig.\,\ref{fig:hmcnc_dist} shows the expected distance measurement from {\it LISA} for HM\,Cnc. We expect an uncertainty of about 1.5 - 2\,kpc, which is significantly better than any current estimates from EM observations (cf. Section~\ref{sec:uncertain_parllax_systems}).

\section{Summary and Conclusions}

In this work we derive updated distances and kinematics for 55 verification binary candidates using parallaxes and proper motions from {\it Gaia} DR3. Using these distances and system properties we calculate the detectability for each source after 1, 3, 6 and 48 months of {\it LISA} observations. We find that 18 verification binaries can be detected after 3 months of {\it LISA} observations and used for science verification. An additional 22 sources will be detected after 48 months of {\it LISA} observations totaling the number of detectable {\it LISA} to a total of 40 sources. The sources consist of 21 DWDs, 16 AM\,CVn binaries, 2 hot subdwarf binaries and 1 ultracompact X-ray binary. In particular AM\,CVn and HP\,Lib are verification binaries with parallax errors below 1\% making them ideal validation sources. 

The number of detectable {\it LISA} binaries has tripled over the last five years since \citet{kup18}. That is mainly due to increasing number of large scale sky surveys, in particular the ELM survey and ZTF were driving the discoveries over the last few years. However, even with the large increase in sources, the sample is still strongly biased towards luminous binaries and sources located in the Northern hemisphere. However, we predict that the number of systems will continue to increase significantly over the next few years with additional large scale surveys coming online over the next few years and strategies need to be developed to perform efficient follow-up for each source before {\it LISA} launches. 

For sources without a measured $\dot{f}$, generally the distance is required to measure the chirp mass. The error on the parallax scales linearly with the error on the chirp mass. We find that the parallax precision has improved by $20\% - 30\%$ between {\it Gaia} DR2 and DR3 and another $\approx30\%$ improvement is expected for {\it Gaia} DR4. We find that on average for verification binaries the GW amplitude is expected to be measured to $\approx10$\% precision, while the inclination is expected to be determined with $\approx15^\circ$ precision. For detectable binaries these average errors decrease to $\approx50$\% and to $\approx40^\circ$ respectively. At present, $40\%$ of the {\it LISA} sources have no measured inclination from EM observations. These are mainly non-eclipsing sources where it is non-trivial to measure an inclination angle and it might be that no progress will be made on the inclination before {\it LISA}. Therefore, even an uncertain inclination measurement from {\it LISA} will be extremely valuable. 

Properties for all sources are collected for this publication is publicly available on the {\it LISA} Consortium GitLab repository   \url{https://gitlab.in2p3.fr/LISA/lisa-verification-binaries}. We wish to keep this list up-to-date for the Consortium and more broader community. Thus we welcome submission requests for new binaries and/or other suggestions.

%% Also note that the akcnowlodgment environment does not support long amounts of text. If you have a lot of people and institutions to acknowledge, do not use this command. Instead, create a new \section{Acknowledgments}.

\begin{acknowledgments}
%\acknowlegments

While working on this study, we were deeply saddened by the loss of Professor Tom Marsh, a world-leading expert on compact binary star systems and a visionary in recognizing the potential of the {\it LISA} mission for the study of these binaries. Professor Marsh's profound knowledge and pioneering insights were invaluable to this study. His enthusiasm and dedication to the field were not only inspiring but also instrumental in shaping the direction of our work.

TK acknowledges support from the National Science Foundation through grant AST \#2107982, from NASA through grant 80NSSC22K0338 and from STScI through grant HST-GO-16659.002-A. Co-funded by the European Union (ERC, CompactBINARIES, 101078773). Views and opinions expressed are however those of the author(s) only and do not necessarily reflect those of the European Union or the European Research Council. Neither the European Union nor the granting authority can be held responsible for them.
VK acknowledges support from the Netherlands Research Council NWO (Rubicon 019.183EN.015 grant).
PJG is partially supported by NRF SARChI grant 111692. Armagh Observatory \& Planetarium is core funded by the Northern Ireland Executive through the Dept for Communities. SS acknowledges support from the DLR grant number / F\"orderkennzeichen: 50OQ1801.

This work presents results from the European Space Agency (ESA) space mission Gaia. Gaia data are being processed by the Gaia Data Processing and Analysis Consortium (DPAC). Funding for the DPAC is provided by national institutions, in particular the institutions participating in the Gaia MultiLateral Agreement (MLA). The Gaia mission website is \url{https://www.cosmos.esa.int/gaia}. The Gaia archive website is \url{https://archives.esac.esa.int/gaia}

\end{acknowledgments}


\vspace{5mm}
\facilities{Gaia}


\software{\texttt{Matplotlib} \citep{hun07}, \texttt{Astropy} \citep{astpy13, astpy18}, \texttt{Numpy} \citep{numpy},  \texttt{LDASOFT} \citep{PhysRevD.101.123021}
       }




\bibliography{refs,biblio}{}
\bibliographystyle{aasjournal}



%\appendix

%\section{Distance estimate for HM Cnc}

\end{document}